\documentclass[aps,prb,twocolumn,showpacs,final,a4paper,superscriptaddress]{revtex4-2}

\usepackage{float}
\usepackage[dvips,final]{graphicx}
\usepackage{dcolumn}
\usepackage{bm}
\usepackage{textcomp}
\usepackage{amsmath,amssymb,amsfonts}
\usepackage{color}
\usepackage{hyperref} 
\hypersetup{colorlinks,linkcolor={blue},citecolor={blue},urlcolor={blue}}

\begin{document}

\title{A Unified Simulation Framework for Correlated Driven-Dissipative Quantum Dynamics}

\author{Thomas Blommel}
\affiliation{Department of Chemistry, University of California, Santa Barbara, California, USA}
\affiliation{Materials Department, University of California, Santa Barbara, California, USA}

\author{Enrico Perfetto}
\affiliation{Dipartimento di Fisica, Universit\`{a} di Roma Tor Vergata, Via della Ricerca Scientifica 1, 00133 Rome, Italy}
\affiliation{INFN, Sezione di Roma Tor Vergata, Via della Ricerca Scientifica 1, 00133 Rome, Italy}

\author{Gianluca Stefanucci}
\affiliation{Dipartimento di Fisica, Universit\`{a} di Roma Tor Vergata, Via della Ricerca Scientifica 1, 00133 Rome, Italy}
\affiliation{INFN, Sezione di Roma Tor Vergata, Via della Ricerca Scientifica 1, 00133 Rome, Italy}

\author{Vojt\v{e}ch Vl\v{c}ek}
\affiliation{Department of Chemistry, University of California, Santa Barbara, California, USA}
\affiliation{Materials Department, University of California, Santa Barbara, California, USA}

\date{\today}%
\begin{abstract}
Time-resolved photoemission spectroscopy provides a unique and direct way to explore the real-time nonequilibrium dynamics of electrons and holes. The formal theory of the spectral function evolution requires inclusion of electronic correlations and dissipation, which are challenging due to the associated long simulation timescales which translate to a high computational cost.  Recent methodological developments, namely the Real-Time Dyson Expansion, as well as theoretical developments of many-body perturbation theory for dissipative systems, have allowed for the study of driven-dissipative interacting quantum systems.  In this work, we implement the hitherto unrealized study of driven-dissipative interacting quantum systems which includes driven dynamical correlations and utilizes these new methods and perturbative expansions.  We illustrate the combined formalism on a prototypical two-band semiconductor model with long-range density-density Coulomb interactions.  We show that the intraband thermalization of conduction band electrons induces nontrivial time-dependent changes in the system's bandstructure and a time-evolving band-gap renormalization (with a reduction by up to 10\%).  We show that the qualitative features are preserved for a variety of parameters, discuss the corresponding spectral dynamics, and provide an outlook on the newly introduced simulation framework, which enables treating electron-electron scattering and dissipation effects on equal footing.
\end{abstract}

\maketitle

\section{Introduction}

Driven and non-equilibrium quantum systems have increasingly 
attracted interest as they enable experimental route to probe quantum 
interactions and energy flow and dissipation at distinct timescales 
(e.g., electron-electron and electron-phonon), addressed by 
pump-probe techniques with increasingly refined time resolution. 
The interest has been further fueled by the remarkable 
practical applications in the fields of quantum computing, 
simulation, and sensing, where the ability to interact with 
external environments can enable novel behaviors.  Systems of particular interest have included optical 
cavities~\cite{Baumann2010,Ritsch2013}, Floquet 
states~\cite{Mori2023,Sato_2020}, exciton-polariton 
dynamics~\cite{Wouters2007,Deng2010,Byrnes2014,Kasprzak2006}, 
entanglement~\cite{Krauter2011,Kastoryano2011}, 
qubits~\cite{DelRe2020,DelRe2024,Rost2025}, and exciton-phonon couplings in solids~\cite{guo2025phononassistedradiativelifetimesexciton,simoni2025firstprinciplesopenquantumdynamics}. For these application as 
well as for use in ultrafast electronics driven by coupling to 
optical fields, it is critical to address the thermalization of 
photoexcited charge carriers, leading to band-gap 
renormalization~\cite{reeves2025,Chernikov_2015_2,Ulstrop_2016,Roth_2019,Kang_2017}. 
Finally, the condensed matter systems under strong external fields 
represent new platforms to study emergent quantum interactions, e.g., 
they enable real time study of exciton formation and their 
scattering.  Despite the availability of  ultrafast experimental 
data, such as that from TR-ARPES~\cite{Boschini2024}, their 
understanding requires, in practice,  theoretical insights to 
interpret and disentangle the precise interactions between charge 
carriers and dissipative baths that give rise to the rich phenomena 
observed.                                   

Despite being of wide technological importance, the driven open quantum systems have, to this point, been 
poorly understood, especially when dynamical correlations arising 
from charge carrier interactions become important.  The reasons for 
this is 
twofold:  \textit{First}, only recently has the nonequilibrium Green's 
function (NEGF) formalism~\cite{SvL2013} been extended to include 
dissipative Lindblad dynamics, enabling an accurate and systematically improvable description 
of system-environment interactions in a 
non-unitary, yet physically consistent, framework.  These extension 
of NEGF theory to open quantum systems has come in both the 
path-integral~\cite{Sieberer_2016,sieberer2023,Fogedby2022,Thompson2023} 
and second quantization approaches~\cite{Stefanucci2024}.  
In both cases, the presence of Lindblad operators leads to new terms appearing in the equations of motion 
for the NEGF, known as the Kadanoff-Baym equations (KBE), as well as 
a reformulation of Many-Body Perturbation Theory 
(MBPT)~\cite{Stefanucci2024}. Here, the two-particle Lindblad and 
quartic Coulomb interaction terms are treated on the same footing.  
\textit{Second}, the timescales associated with dissipative dynamics are 
typically long, especially in the case of thermalization effects, and they have been out of reach for numerical integration routines. This is due to the computational cost of KBE, which scales cubically with the final integration time.  This cost comes from 
the presence of expensive history integrals in the KBE, and much work 
has been done to develop efficient 
integration~\cite{Truncation,AcceleratedCollapse,KayeComp2021,EB_DynamicsGKBA,meirinhos2022adaptive,Exciton_GKBA,G1_G2,Pavlyukh2024,Blommel2024,Blommel2025,Reeves2024} 
and extrapolation~\cite{DMD1T,DMD2T,Extension} methods.  One such 
method, the Real-Time Dyson Expansion (RTDE)~\cite{Reeves2024}, is 
specifically well-suited to capture excited-state photoemission 
spectra data obtained from TR-ARPES experiments, as it scales 
linearly with the measurement time, allowing for the first ever study 
of long-time spectral dynamics of interacting systems in the presence 
of thermal baths.  The RTDE goes beyond static and mean-field 
approximations to the self-energy via a bare expansion of 
higher-order diagrams appearing in MBPT. As shown here, RTDE thus includes 
dynamical correlations originating both from Fermion interactions as 
well as the presence of dissipative baths.                                                              

The recently developed RTDE has been used to study band-gap renormalization in 
model semiconductor systems with weak-to-moderate long-range Coulomb 
interactions~\cite{reeves2025}.  The presence of excited charge 
carriers leads to band-gap closure due to attraction between 
conduction band electrons and valence band holes.  In this work, we 
expand on these results by including dissipative terms into the open 
system Hamiltonian, and consequently modifying the RTDE equations of 
motion, allowing for electrons to thermalize after the initial 
driving pulse.  We show that the excited-state band structure has 
non-trivial dependence on the system's kinetic energy, or 
equivalently, the populations locations in the band.  We see that the 
band-gap monotonically shrinks as the electrons thermalize and 
approach the band edge.  These dynamical effects become stronger with increasing inter- and intraband interaction strengths.

The manuscript will first review the RT-DE formalism expanded to include the dissipative (Lindblad) terms. Next we introduce a practical model, which is studied in the results section. The conclusions provide comments on the theory and implementation, as well as an outlook for future expansion of this methodology.

\section{RTDE for dissipative KBE}
\label{sec:method}

\begin{figure}
    \centering
    \centerline{\includegraphics[width=1.5\linewidth]{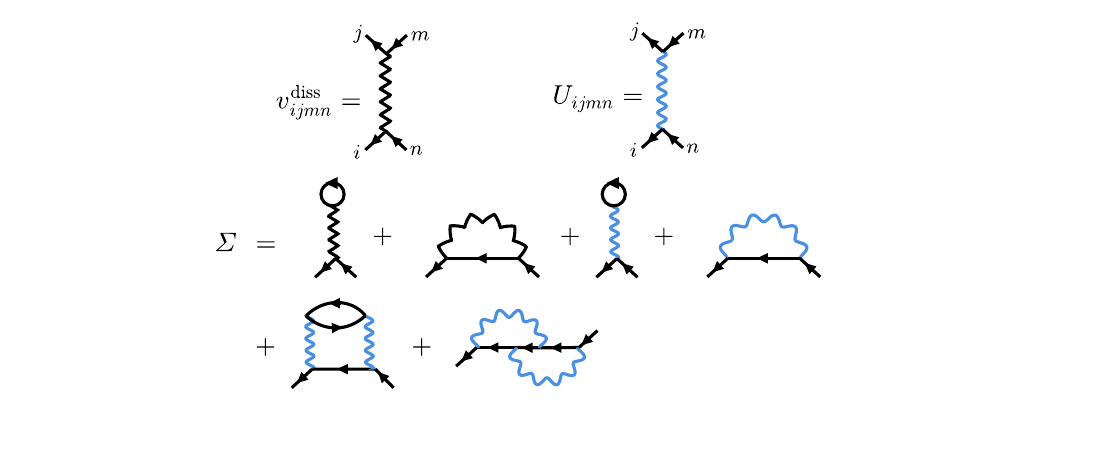}}
    \caption{Self-energy diagrams arising from the MBPT expansion of quartic Lindblad and Coulomb terms in the interacting open-system Hamiltonian.  Second order diagrams involving the dissipative interaction line $v^{\rm{diss}}$ are ignored in this work.}
    \label{fig:Diagrams}
\end{figure}
To study the effects of inter- and intra-band dissipative dynamics on 
the band structure of materials, we calculate the time-resolved 
spectral function as:~\cite{freericks_theoretical_2009,perfetto_first-principles_2016}   
\begin{equation}
    \mathcal{A}^<(\vec{k},\omega,T_m) = \int dtdt' e^{-i\omega(t-t')} \mathcal{S}(t-T_m)\mathcal{S}(t'-T_m) G_{\vec{k}}^<(t,t'), \label{Eq:PEspectrum}
\end{equation}
where the lesser component of the spectral function represents the distribution of electrons around the time $T_m$.
In the expression above, $\mathcal{S}(t-T_m)$ represents the envelope of a probe pulse 
centered at $T_m$, which 
we take to be a Gaussian with width $T_w$.  The non-equilibrium 
Green's function, $G_{\vec{k}}^<(t,t')$, can be obtained through a 
variety of numerical methods.  In this work, we use the Real-Time 
Dyson Expansion (RTDE)~\cite{Reeves2024}, which is an approximation to the KBE.  The RTDE is very efficient at calculating the 
photoemission data, as the computational cost scales as $T_m T_w^2$, 
whereas the KBE scale as $T_m^3$ if solving them na\"{i}vely, or 
$T_m^2 \log T_m$ using efficient compression-based 
integrators. Finally, the total quasiparticle (QP) density of states is obtained from the retarded spectral function $A^R(\vec{k},\omega,T_m)$.             

To model the relaxation of the excited system back to equilibrium, the interacting Hamiltonian, described as
\begin{gather}
\hat{H}(t)=\sum_{mn}h_{mn}(t)\hat{d}^\dagger_m\hat{d}_n+\hat{H}_{\rm 
int}\\
\hat{H}_{\rm 
int} = \frac{}{}\frac{1}{2}\sum_{ijkl}U_{ijkl}(t)\hat{d}^\dagger_i\hat{d}^\dagger_j\hat{d}_k\hat{d}_l,
\end{gather}
is coupled to a zero-temperature reservoir using static (i.e., time-independent) particle-hole 
Lindblad operators  
\begin{align}
    \hat{L}_\gamma(t)&= a^\gamma_{mn}\hat{d}^\dagger_m\hat{d}_n
	\label{lindop}
\end{align}
and hence
\begin{align}
\sum_{\gamma}	\hat{L}^{\dagger}_{\gamma}\hat{L}_{\gamma}=
\sum_{mn}
|a_{mn}|^{2}\hat{d}^{\dagger}_m\hat{d}_n
\hat{d}^{\dagger}_n\hat{d}_m.
\end{align}

Formally, the effects of the Lindblad operators can be included into the KBE via a 
$\Phi$-derivable perturbation theory in a similar manner as the 
interaction expansion for quartic electron-electron interactions in 
the interacting Hamiltonian. The dissipative 
KBE read (in matrix form)~\cite{Stefanucci2024}
\begin{subequations}
\begin{align}
\Big[i\frac{d}{dt}-
h_{o}(t)\Big]G^{<}(t,t')&- 2i\ell^{<}(t)
G^{\rm A}(t,t')
\nonumber\\
&=\big[
\Sigma^{<}\cdot G^{\rm A}+
\Sigma^{\rm R}\cdot G^{<}\big](t,t'),
\label{kb<t}
\end{align}
\begin{align}
G^{<}(t,t')\Big[\frac{1}{i}\frac{\overleftarrow{d}}{dt'}-
h^{\dagger}_{o}(t')
\Big]&- 2iG^{\rm R}(t,t')\ell^{<}(t')
\nonumber\\
&=\big[
G^{<}\cdot \Sigma^{\rm A}+
G^{\rm R} \cdot\Sigma^{<}
\big](t,t'),
\label{kb<tp}
\end{align}
\begin{align}
\Big[i\frac{d}{dt}-h_{o}(t)\Big]G^{\rm R}(t,t')=\delta(t,t')
+\big[
\Sigma^{\rm R} \cdot G^{\rm R}\big](t,t'),
\label{kbrt}
\end{align}
\end{subequations}
and the like for the greater and advanced Green's functions. 
Lindblad dissipation is responsible for a non-hermitian
renormalization of the single-particle Hamiltonian 
\begin{align}
h(t)\to h_{o}(t)&=h(t) - i\big[\ell^>(t) + \ell^<(t)\big] -R(t),
\end{align}
where
\begin{align}
	v^{\mathrm{diss}}_{jimn}(t) &= \sum_\gamma2a^{\gamma*}_{ni}(t)a^{\gamma}_{jm}(t),\\
    l^>_{in}(t) &= 
	\frac{1}{2}\sum_{mj}v^{\mathrm{diss}}_{ijnm}(t)(\delta_{mj}-\rho_{mj}(t)),\\
    l^<_{in}(t) &= 
	\frac{1}{2}\sum_{mj}v^{\mathrm{diss}}_{jimn}(t)\rho_{mj}(t),\\
    R_{in} &= \frac{1}{2}\sum_{mj}(v^{\mathrm{diss}}_{ijmn}(t) - 
	v^{\mathrm{diss}}_{jinm}(t))G^{<}_{mj}(t,t)
\end{align}
as well as for the appearance of the terms $2i\ell^{<}
G^{\rm A}$ and $2iG^{\rm R}\ell^{<}$ in the equations of motion for 
$G^{\lessgtr}$. On the Keldysh contour, the dissipation-induced interaction, $v^{\mathrm{diss}}$, can be combined with the physical Coulomb interaction, $U$; consequently, the  
expansion of the correlation self-energy $\Sigma$ in terms of Feynman 
diagrams with arguments on the Keldysh contour formally retains the same structure.

In this work, we assume the magnitudes of the Lindblad coefficients 
to be much smaller than the energy scales associated with electronic 
kinetic and interaction energies.     
Consequently, we retain only the mean-field contributions in the 
expansion of the dissipative self-energy. In contrast, Coulomb 
correlations are treated within the second-Born (2B) approximation.    

The dissipative KBE with mean-field dissipation and 2B correlations 
are solved using the RTDE method~\cite{Reeves2024}. In RTDE we first 
solve for the one-electron density matrix $\rho^{<}(t)=-iG^{<}(t,t)$ at the mean-field 
level. Setting $\Sigma=0$, subtracting  Eq.~(\ref{kb<tp}) from 
Eq.~(\ref{kb<t}), and evaluating the result in $t=t'$ we find that 
the mean-field $\rho^{<}$ satisfies the 
Lyapunov equation 
\begin{align}
    \frac{d}{dt}\rho^{<,\mathrm{MF}}(t) &= -ih_o(t)\rho^{<,\mathrm{MF}}(t) + i\rho^{<,\mathrm{MF}}(t)h_o^\dagger(t) + 
	2\ell^<(t).
\end{align}
Away from the $t=t'$ diagonal, we solve Eq.~(\ref{kb<t}) for $t>t'$ 
using $G^{<}(t,t)=i\rho^{<,\mathrm{MF}}(t)$ as initial condition, and following the Generalized 
Kadanoff-Baym Ansatz, for the 2B collision integral in the 
right hand side
\begin{align}
G^{\lessgtr}(t,t')\simeq -G^{\rm 
R, MF}(t,t')\rho^{\lessgtr}(t')+\rho^{\lessgtr}(t)
G^{\rm A, MF}(t,t'),\label{eq:Gtt_gkba}
\end{align}
where $\rho^{>}=\rho^{<}-1$, and 
\begin{align}
G^{\rm 
R, MF}(t,t')=[G^{\rm A, MF}(t',t)]^{\dagger}=-i\theta(t-t')Te ^{-i\int_{t'}^{t}d\bar{t}\,h_{o}(\bar{t})}
\end{align}
is the \textit{mean-field} retarded Green's function, see Eq.~(\ref{kbrt}).
Note that the open system $G^{R}$ has the same structure as  
that used in the GKBA treatment of time-dependent 
quantum transport~\cite{latini_charge_2014,tuovinen_time-linear_2023}. 
Finally, following the development in Ref.~\cite{Reeves2024}, 
one can show that the integro-differential equation for the dynamically correlated $G^{R}(t,t')$ is
equivalent to the following coupled system of ordinary 
differential equations, in which $G^{\lessgtr}(t,t) = i\rho^{\lessgtr,MF}(t)$ following Eq.~(\ref{eq:Gtt_gkba})
\begin{widetext}
\begin{subequations}
\begin{align}
    \partial_t G^R_{ij}(t,t') &= -i[\sum_xh^{\text{MF}}_{o,ix}(t)G^R_{xj}(t,t') - \tilde{I}^R_{ij}(t,t')],\\
    \tilde{I}^R_{ij}(t,t') &= - i \sum_{lk}U_{ilkq}(t)\mathcal{F}_{qklj}(t,t'),\\
    \partial_t \mathcal{F}_{qklj}(t,t') &=- \sum_{mnpy}\left[U_{mnpy}(t)-U_{nmpy}(t)\right]\left[\rho_{qm}^{>, \mathrm{MF}}(t) \rho_{kn}^{>, \mathrm{MF}}(t) \rho_{pl}^{<, \mathrm{MF}}(t)-\rho_{qm}^{<,\mathrm{MF}}(t) \rho_{kn}^{<, \mathrm{MF}}(t) \rho_{pl}^{>, \mathrm{MF}}(t)\right] G_{yj}^{\mathrm{R}}\left(t, t^{\prime}\right)\label{Eq:FR}\\
    &\quad\quad\quad -i\sum_x\left[h_{o,qx}^{\text{MF}}(t)\mathcal{F}_{xklj}(t,t') + h_{o,kx}^{\text{MF}}(t)\mathcal{F}_{qxlj}(t,t') - \mathcal{F}_{qkxj}(t,t')h_{o,lx}^{\text{MF},*}(t)\right]\nonumber.
\end{align}
\end{subequations}
\end{widetext}
to be solved with boundary conditions $G^{R}(t,t)=-i$ and $\mathcal{F}(t,t)=0$.  The equations for the lesser component are nearly identical and can be obtained with the replacement $G^R\rightarrow G^<$, removing the $\rho^>\rho^>\rho^<$ term in Eq.~\ref{Eq:FR}, and removing the (-) sign in front of the $\rho^<\rho^<\rho^>$ term.  The boundary condition for the lesser component Green's function is $G^{<}(t,t)=i\rho^{<}(t)$, and $\mathcal{F}(t,t)\approx0$ has been shown to yield accurate results, as discussed in Ref.~\cite{Reeves2024}.
	
\section{Two-band dissipative model}   
\label{sec:model}

To study the deformation of bands due to the dissipative movement of excited electrons and holes we use a periodic extended Hubbard Hamiltonian with two bands
\begin{align}
    \hat{H}(t) &= 
	\sum_{\substack{\alpha,\beta\\i,j,\sigma}}h^{\alpha\beta}_{ij}(t)\hat{d}^{\dagger\alpha}_{i,\sigma}\hat{d}^{\beta}_{j,\sigma}\\
    &+ U_\mathrm{intra}\Bigg{[}\sum_{i,\alpha}\hat{n}^\alpha_{i\uparrow}\hat{n}^{\alpha}_{i\downarrow} +\sum_{i < j,\alpha} \frac{\hat{n}_{i}^\alpha \hat{n}_{j}^\alpha}{\varepsilon|\vec{r}_i - \vec{r}_j|}\Bigg{]}\nonumber\\
    &+U_\mathrm{inter}\Bigg{[}\sum_{i}\hat{n}_{i}^c\hat{n}_{i}^v + \sum_{i < j} \frac{\hat{n}_{i}^c \hat{n}_{j}^v}{\varepsilon|\vec{r}_i - \vec{r}_j|}\Bigg{]}\nonumber\\
    h(t) &= h^{(0)} + h^{\text{laser}}(t)     \label{Eq:H}
\\
    h^{(0),\alpha\beta}_{ij} &= J_\alpha\delta_{\alpha\beta}\delta_{\langle i,j\rangle}+ \epsilon_\alpha\delta_{\alpha\beta}\nonumber\\
    h^{\mathrm{laser},\alpha\beta}_{ij}(t) &= \delta_{ij}(1-\delta_{\alpha\beta})
        E\cos(\omega_l(t-t_0)) \mathrm{e}^{-\frac{(t-t_0)^2}{2T_l^2}}\nonumber.
\end{align}
In this work, we use measure units relative to the intraband nearest-neighbor hopping, $J_c = J_v = 1$.  The on-site potentials, $\epsilon_\alpha$ are tuned such that the bandgap of the equilibrium system is $E_\text{gap}=5$.  $U_\text{inter}$ and $U_\text{inter}$ determine the strength of the density-density Coulomb interactions in the system, with the long-range interactions screened by the dielectric constant $\varepsilon=5$.  We work in the moderate interaction strength regime, where $U_{\text{intra}}=1,3$ and $U_{\text{inter}}=0,0.5,1$.

In order to achieve sufficient momentum resolution of the electronic dynamics, we choose the number of sites to be $N_s=16$ and restrict our calculations to one dimension.  The width of the probe window, $T_w$, balances the trade-off between spectral and time resolution.  We have found that $T_w = 8$ provides us with a good balance between the two.  All of our calculations use fourth order Runge-Kutta to integrate the equations of motion.  We use a timestep of $h=0.025$ for the Hartree-Fock Density Matrix integration, and $h=0.05$ for the Green's functions away from the $t=t'$ diagonal.

To model the dissipation of energy into a zero-temperature thermal reservoir, we choose the Lindblad coefficients, $a_{(\vec{k},\alpha)(\vec{k'},\beta)}$, such that the scattering processes they describe always result in the creation of an electron with lower single-particle energy than the one which was destroyed.  The simplest model of this process is to only allow intraband transitions between states that are nearest neighbors in momentum space
\begin{equation}
    a^{\text{intra}}_{(\vec{k},\alpha)(\vec{k'},\alpha)} = \begin{cases}
        \sqrt{\Gamma^\text{intra}} & E(\vec{k}) < E(\vec{k'})\;\land \;|\vec{k}-\vec{k'}| = \frac{2\pi}{N_s}\\
        0 & \text{otherwise}
    \end{cases}
    \label{Eq:Gamma_intra}
\end{equation}
where $\Gamma^\text{intra}$ is a dissipation rate parameter we set to $0.003$, much smaller than the energy scale of the hopping parameter, $J_v=1$.  In the presence of electron interactions, these choices of Lindblad parameters may not minimize the total energy of the system, as they are chosen only with knowledge of the single-particle Hamiltonian.  This becomes an issue when the inter-particle interaction strengths become comparable to the kinetic energy scales, however this will happen when the diagrammatic perturbation theory that RTDE is based on will begin to fail as well.  It is also possible to allow for thermal distributions of electrons within a band by allowing processes that increase the energy of the system, however we neglect these for the purposes of this analysis.

The particle-hole loss operators we use can also model radiative recombination effects between the conduction and valence band.  We model these processes by only allowing transitions which conserve momentum, and exponentially damp radiative processes with the energy difference between bands at a given $\vec{k}$ point
\begin{equation}
    a^{\text{inter}}_{(\vec{k},v)(\vec{k},c)} = 
        \sqrt{\Gamma^\text{inter}}e^{-\frac{1}{2}(E_c(\vec{k}) - E_v(\vec{k}) - E_g)}.
        \label{Eq:Gamma_inter}
\end{equation}

\begin{figure}
    \centering
    \includegraphics[width=\linewidth]{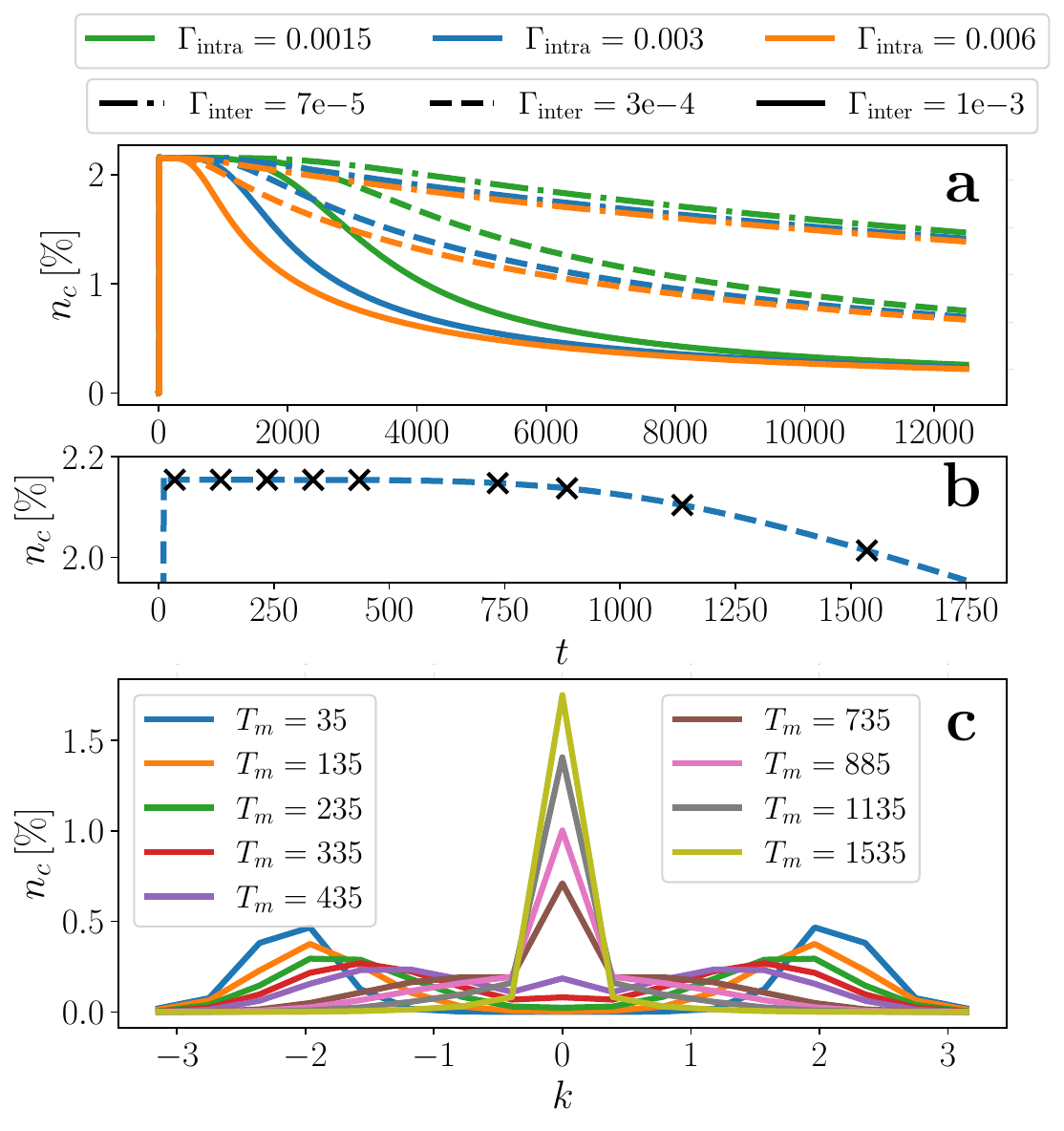}
    \caption{Dynamics of the density matrix under the influence of Lindblad dissipation.  (a) Conduction band occupation for several parameters of $\Gamma_\mathrm{intra}$ and $\Gamma_\mathrm{inter}$  (b) Conduction band occupation for the parameters chosen for the rest of the paper.  Black $\times$ marks the photoemission spectrum measurement times, and correspond to the distributions shown in (c). (c) $k$-resolved densities at photoemission measurement times $T_m$.  Data is obtained from calculations with parameters $U_{\text{intra}}=3$, and $U_{\text{inter}}=0.5$.}
    \label{Fig:rho_dynamics}
\end{figure}

\section{Results}
\begin{figure}[h!]
    \centering\includegraphics[width=\linewidth]{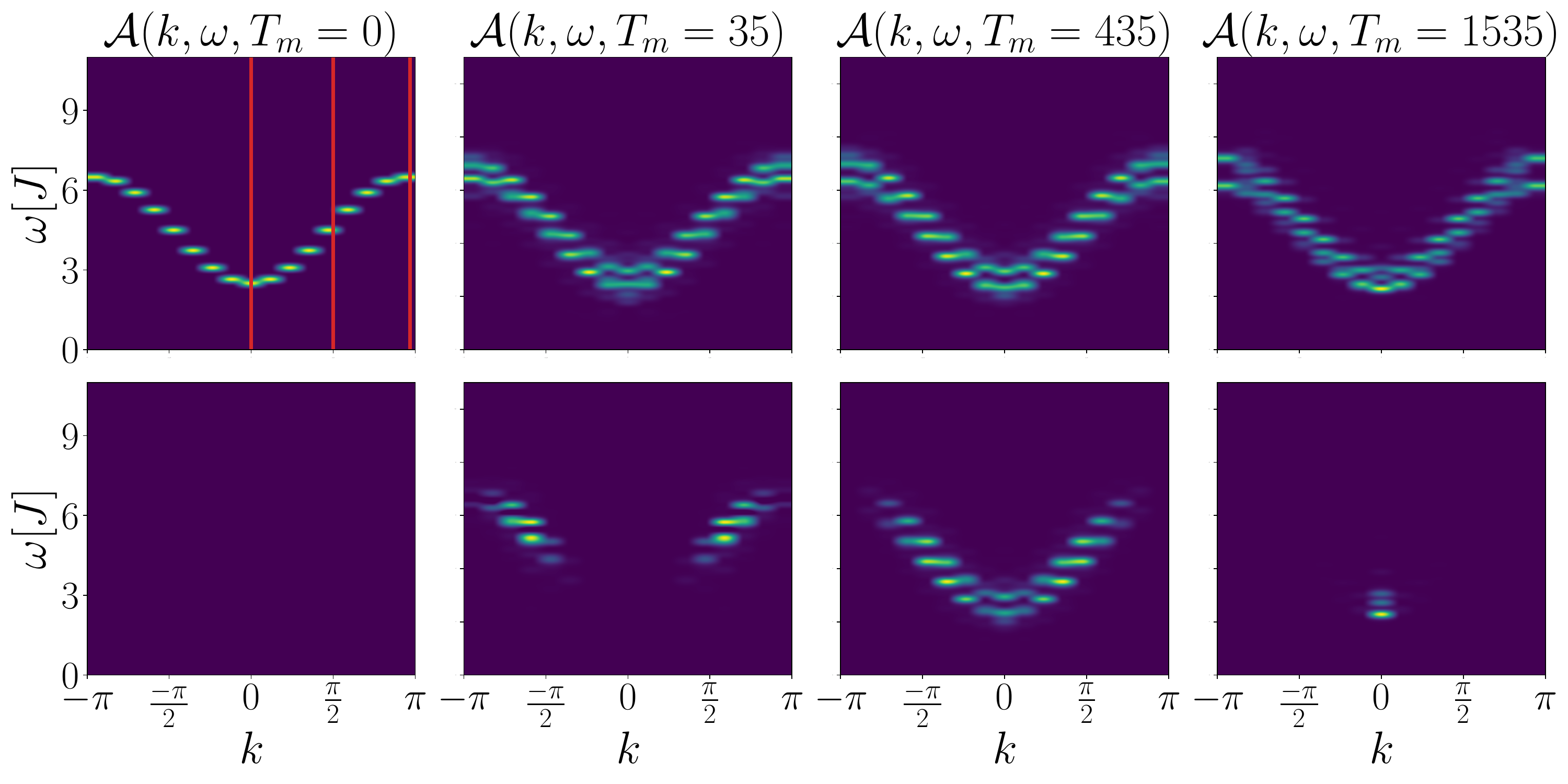}
    \caption{Retarded (top row) and Lesser (bottom row) photoemission spectrum calculated from RTDE.  Parameters for these calculations are $U_{\text{intra}}=3$, and $U_{\text{inter}}=0.5$. Red bars show the $k$-points plotted in Fig.~\ref{Fig:spectrum_splitting}.}
    \label{Fig:U3_05_spectra}
\end{figure}

\label{sec:density_matrix}
In this section, we discuss the dynamics of the photoemission spectrum as the system thermalizes under the influence of the Lindblad operators.  We begin this analysis by first looking at the momentum-resolved density matrix and observe the dissipation of energy as the electrons accumulate at $k=\Gamma$.  We next choose a set of measurement times that span the timescale of this thermalization process and evaluate the photoemission spectrum.

\subsection{Density Matrix Dynamics}
The system is excited at time $t_0=10$ using a semi-classical dipole coupling to an electric field, Eq.~\ref{Eq:H}.  Since we wish to study how the excited-state band structure changes as electrons dissipate energy, we first excite high into the band at $k=\pm \frac{2\pi}{3}$.  To accomplish this, we tune the frequency of the excitation field to the energy difference between the equilibrium conduction and valence bands at those $k$-points, which is $\omega_l=11$.  We set the pulse width to  $T_l=0.7$.  We tune the laser amplitude such that we achieve an excited population of $n_c\approx2.1\%$, which is in the range of experimentally achievable populations for photoexcited semiconductors~\cite{Hedayat_2021,Spataru_2004,CALLAN2000167}.  Despite the form of our intraband dissipation operator, Eq.~\ref{Eq:Gamma_intra}, appearing to break translational invariance, the symmetry is preserved by $\sum_\gamma \hat{L}^\dagger_\gamma \hat{L}_\gamma$.  We have checked that our integration routine preserves translational invariance to within machine precision.

Due to the relaxation timescales seen observed in pump-probe spectroscopy experiments, we expect that radiative recombination processes take place on timescales which are much longer than the intraband relaxation of the excited electrons.  We also expect that radiative processes very rarely occur away from $k=\Gamma$, so the population of the conduction band will stay almost constant while the excited electrons are high in the band.  We observe this behavior in  Fig.~\ref{Fig:rho_dynamics}(a), where we present data for nine combinations of $\Gamma_\mathrm{inter}$ and $\Gamma_\mathrm{intra}$, where $\Gamma_\mathrm{intra}$ is greater than $\Gamma_\mathrm{inter}$.  The plateau of the conduction band electrons at short times is due to our driving field being tuned to achieve a photodoped population of charge carriers high in the band, far from $k=\Gamma$.  The length of this plateau becomes larger as $\Gamma_\mathrm{intra}$ decreases due to the longer thermalization timescales, meaning electrons take longer to begin accumulating at $k=\Gamma$.  It is also clear that once all conduction electrons have thermalized and sit at $k=\Gamma$, the radiative rate is set entirely by $\Gamma_\mathrm{inter}$, and the depopulation curves differ mostly by the time shift arising from the different intraband thermalization timescales. 

Fig.~\ref{Fig:rho_dynamics}(b) shows the trajectory of the conduction band populations for the choices of $\Gamma_\mathrm{inter}$ and $\Gamma_\mathrm{intra}$ we will use for the remainder of our analysis, which corresponds to the middle value for both parameters shown in (a).  We place black $\times$ marks in (b) to denote the times, $T_m$, where we measure the photoemission spectrum of the system.  Note the difference in timescales between (a) and (b); we only make measurements of the spectrum during the intraband thermalization process.  As the system undergoes the radiative process, the photoemission spectrum will slowly and monotonically approach the equilibrium band structure, giving less interesting dynamics due to the fact that the charge carrier momentum distribution will be unchanging.

The $k$-resolved conduction band populations are shown in (c), where the times $T_m$ correspond to the black $\times$ marks in (b).  We see that very shortly after the pulse, at $T_m=35$, the populations are centered around $k=\pm\frac{2\pi}{3}$, which exactly corresponds to the gap in the equilibrium band structure at the driving frequency $\omega=11$.  This is consistent with our choice of system-light coupling Hamiltonian and driving field frequency discussed in the beginning of this section.  The finite width of the $k$-resolved density distribution comes from the finite width of the driving laser field.  The dynamics of the conduction band electrons are consistent with our expectations of the populations slowly losing energy and converging at $k=\Gamma$, and the zero-temperature nature of the bath ensures that no electrons move back up the band.  Despite the coupling to a dissipative bath, our choice of Lindblad operators conserves the number of charge carriers in the system, which has been verified numerically within machine precision.  The accumulation of electrons at $k=\Gamma$ may become energetically unfavorable in situations where the interaction energy becomes large relative to the difference in single-particle energies at each $k$-point.  The inability of our model to capture these scenarios is due to the simple form of the Lindblad coefficients we have chosen.  In situations where ab-initio calculations are required, these coefficients can become more robust and contain information about phonon fields, or may contain feedback mechanisms to self-consistently modify system-bath couplings based on excited-state system properties.  In the model system studied in this work, we find the simple form of our dissipative operators to be adequate in describing the thermalization processes we wish to capture, and a monotonic decay of total energy.

\subsection{Spectral Dynamics}
\begin{figure}[h!]
    \centering
    \includegraphics[width=\linewidth]{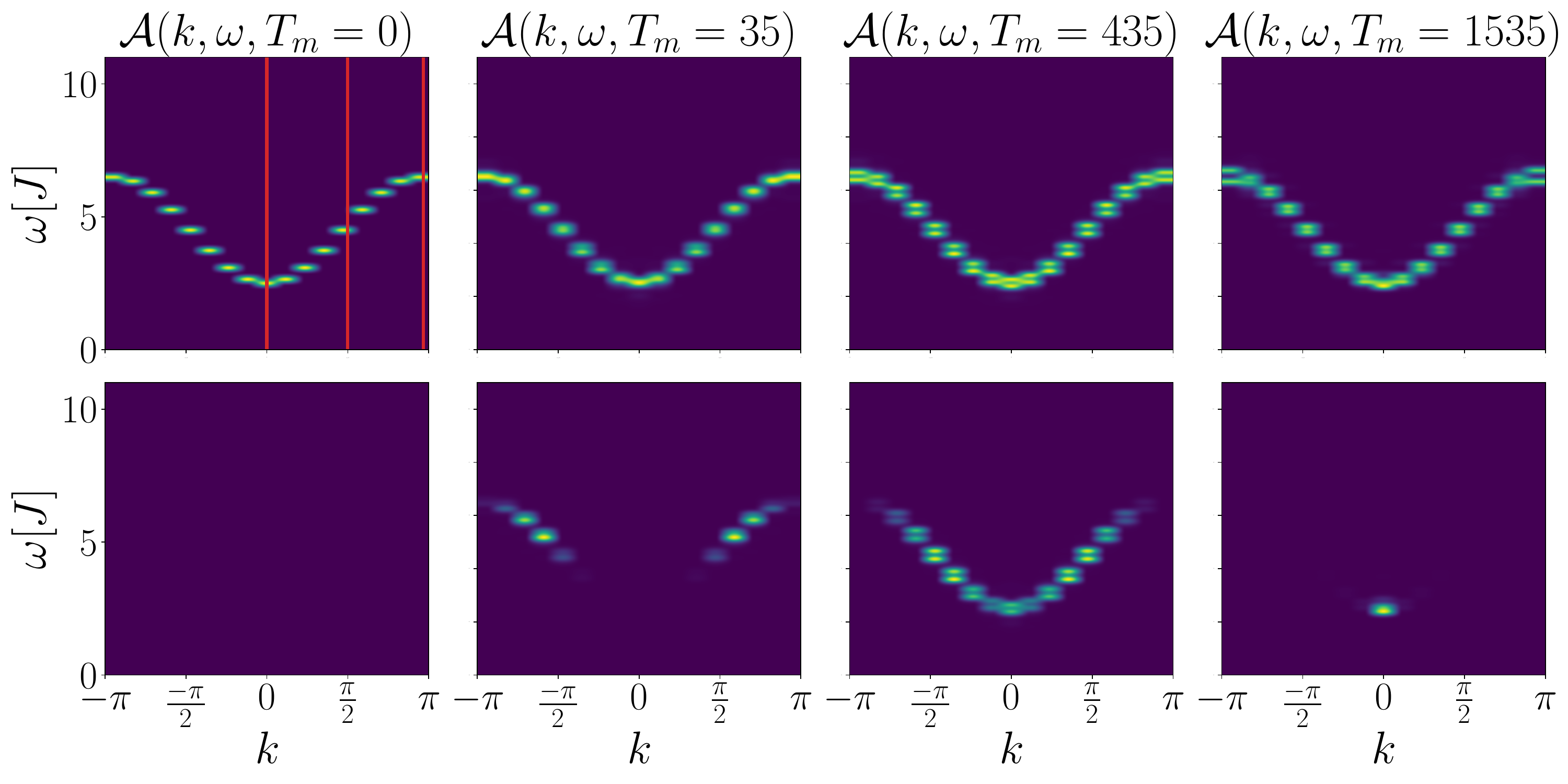}
    \caption{Retarded (top row) and Lesser (bottom row) photoemission spectrum calculated from RTDE.  Parameters for these calculations are $U_{\text{intra}}=1$, and $U_{\text{inter}}=0.5$. Red bars show the $k$-points plotted in Fig.~\ref{Fig:spectrum_splitting}.}
    \label{Fig:U1_05_spectra}
\end{figure}

Using RTDE, we are able to evaluate the photoemission spectrum where the probe field is centered at the $T_m$ from Fig.~\ref{Fig:rho_dynamics}, where the measurement times span the timescale of intraband thermalization in the excited system. These spectra are shown in Fig.~\ref{Fig:U3_05_spectra}, where the bottom row shows the lesser component, which is measurable in real systems through TR-ARPES experiments.  This lesser component contains information about the electronic populations, specifically we can extract the conduction band populations from $n_c(\vec{k},T_m) = \int_\mu^\infty d\omega A^<(\vec{k},\omega, T_m)$.  In the top row, we show the spectrum computed from the Retarded component so that we can visualize the dynamics of the entire band structure.  We show only the conduction band dynamics, as particle-hole symmetry gives the valence band via mirroring across $\omega=0$. 

The most obvious feature of these excited spectra is the splitting of the conduction band at each $k$-point.  When driven out of equilibrium, interaction-induced correlations due to charge fluctuations causes each band to undergo splitting, which resembles in nature the Mott transition in equilibrium.  These sub-bands split in such a way that several poles move into the gap, and several away from the gap.  This movement of spectral weight into the gap causes the bandgap of the system to shrink relative to the equilibrium bandgap, and several smaller satellites appear deeper into the gap.  This shrinking of the bandgap is consistent with other numerical~\cite{reeves2025} and experimental~\cite{Chernikov_2015_2,Ulstrop_2016,Roth_2019,Kang_2017} results.   We recall that RTDE is built upon MBPT and in this work we use the Second Born approximation, so it is important to understand if the observed dynamics persist when the interactions are weaker, and are not just an artifact of the self-energy approximation.  To do so, we present the photoemission spectrum for $U_{\text{intra}}=1$ and $U_{\text{inter}}=0.5$ in Fig.~\ref{Fig:U1_05_spectra}. Compared to the data with larger $U$, these spectra have decreased splitting amplitudes and smaller number of QP peaks and satellites.  However the general trend of the dynamics of the excited state bands are clearly similar to the results from the more strongly interacting system.

\begin{figure}[h!]
    \centering
    \includegraphics[width=\linewidth]{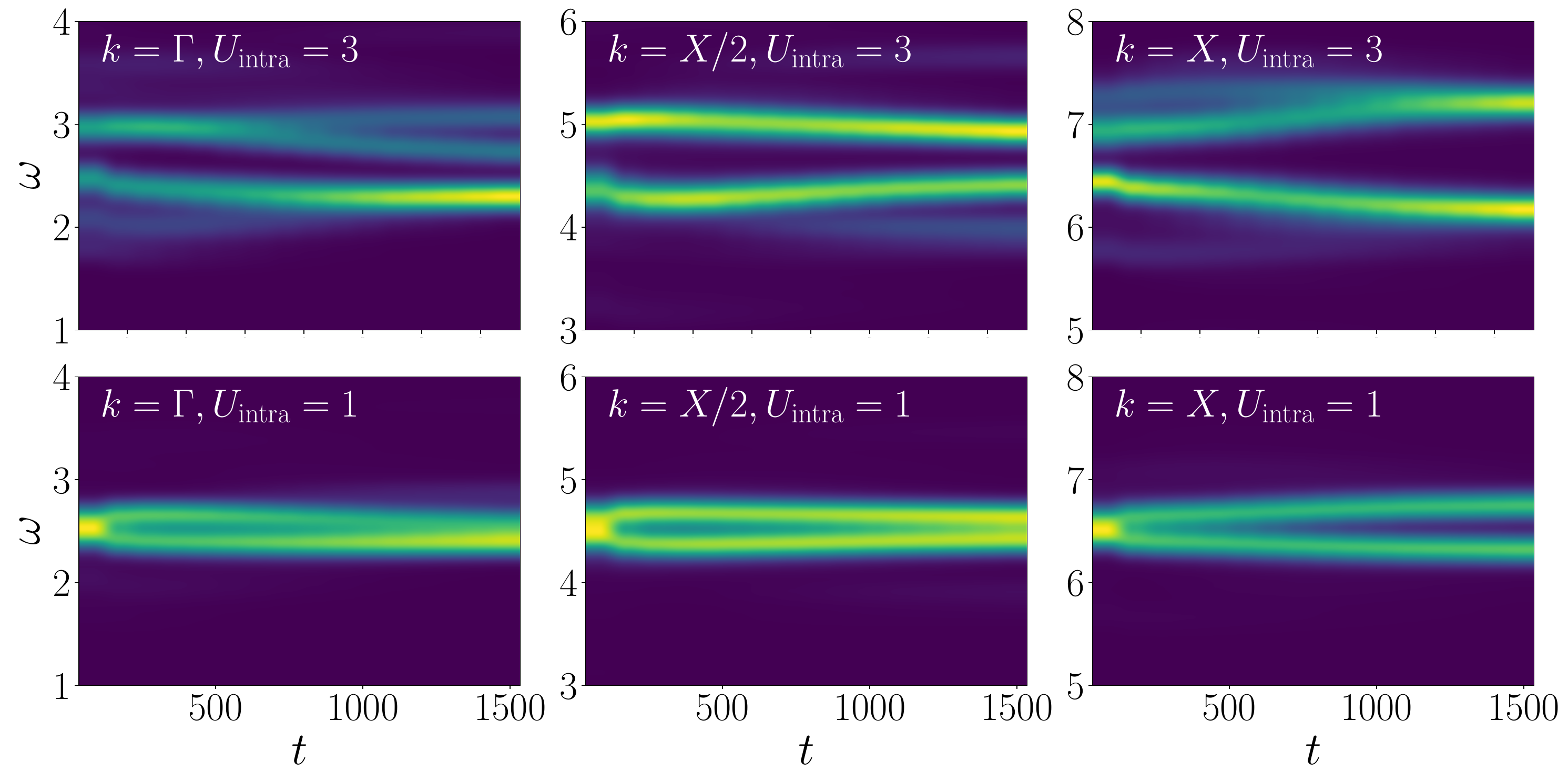}
    \caption{Time-dependent spectra at three different points in the Brillouin Zone.  Top row: $U_{\text{intra}}=3$, and $U_{\text{inter}}=0.5$. Bottom row: $U_{\text{intra}}=1$, and $U_{\text{inter}}=0.5$.  The y-axis on each paned is centered around the equilibrium position of the conduction band at the given $k$-point.}
    \label{Fig:spectrum_splitting}
\end{figure}

From Figs.~\ref{Fig:U3_05_spectra}-\ref{Fig:U1_05_spectra}, it is clear that the spectral function has non-trivial dynamics as the system thermalizes.  These changes to the non-equilibrium bands are despite the fact that the number of conduction electrons stays mostly constant, as there has not been significant radiative recombination by time $T_m=1535$.  This illustrates the importance of the (long-range) $k$-dependence of the electronic interaction tensor in shaping the non-equilibrium band structure.  In order to more closely analyze the dynamics of the band structure, we plot slices of the time dependent spectra, $A(\vec{k}_0,\omega,t)$ at $\vec{k}_0=\Gamma,X/2,X$ in Fig.~\ref{Fig:spectrum_splitting}.  The three $k$ points chosen correspond to the red bars in Figs.~\ref{Fig:U3_05_spectra}{-}\ref{Fig:U1_05_spectra}.  

First, we focus on the dynamics at $k=\Gamma$.  Immediately after the excitation, the conduction band splits into two main QP peaks for both $U_\text{intra}=1$ and $3$, and several lower intensity satellites appear farther away from the two QP peaks for $U_\text{intra}=3$. This is consistent with the shake-up satellites observed in equilibrium spectral functions, but here they are a combined effect due to the pre-excitation by the initial light pulse. The satellite spectral signatures are short lived -- they eventually recombine with the main two QP peaks present in the non-equilibrium spectrum.  The QP peak that splits inwards towards the gap continues to move inwards and further close the gap.  This implies that the attraction between conduction electrons and valence holes becomes stronger as the system thermalizes and both species move towards $\Gamma$.  

\begin{figure}
    \centering
    \includegraphics[width=\linewidth]{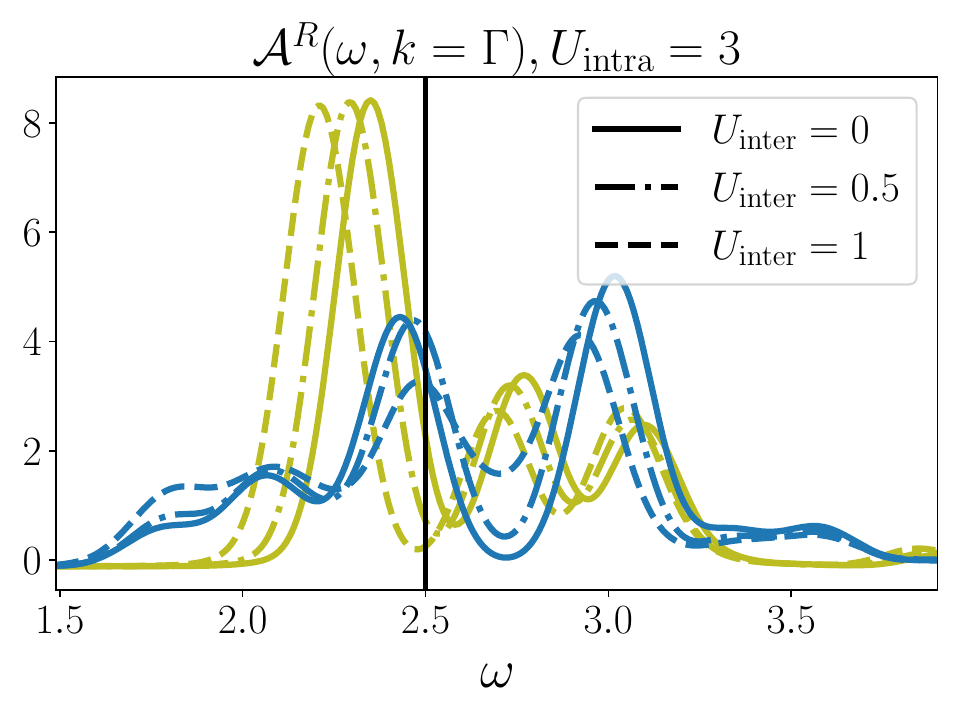}
    \caption{Retarded conduction band spectrum at $k=\Gamma$ for several values of $U_\text{inter}$.  Blue lines at $T_m=35$ and gold lines at $T_m=1535$.}
    \label{Fig:inter}
\end{figure}

For $U_\text{intra}=3$, the gap between the conduction and valance QP peaks shrinks by $1.3\%$ immediately after the excitation when the conduction electrons sit at $k=\pm\frac{2\pi}{3}$.  After the intraband thermalization has occurred, and all the conduction electrons are located at $\Gamma$, the renormalization of the QP bandgap has increased to $8.4\%$.  These percentages decrease as we decrease the intraband interaction to $U_\text{intra}=1$. For the latter parameter, the gap actually increases by $1\%$ immediately after the excitation, whereas the gap is is gradually renormalized by up to $3.6\%$; the latter value is observed in long times after thermalization when the excited charge carriers are located mostly at $\Gamma$.  When estimating the percent renormalization of the gap shortly after the pulse, we only consider the location of the QP peak nearest to the chemical potential, i.e., we ignore the ``in-gap'' satellite features.  These satellites are due to the large intraband interactions and do not originate from the attraction between conduction and valence charge carriers.  Fig.~\ref{Fig:inter} shows that at $k=\Gamma$, these satellites persist even in the limit where $U_\text{inter}=0$.  Furthermore, for $U_\text{inter}=0$, the QP peak is renormalized by $3.3\%$ between times $T_m=35$ and 1535, but increasing the interband interaction to $U_\text{inter}=1$ causes this renormalization to $10.5\%$, showing that the interband interactions play the majority of the role in the shrinking of the gap.  The $k$-dependence of the interband interaction tensor causes the effective interaction between conduction electrons and valence holes to increase as they thermalize towards $\Gamma$.

Fig.~\ref{Fig:spectrum_splitting} also shows us that the dynamics of the conduction band is not uniform across every $k$-point, and that the evolution of the electronic states for distinct intraband interaction strengths, $U_\text{intra}=3$ and $U_\text{intra}=1$, are qualitatively similar.  At $k=X$, the inner peak shifts more than at $k=\Gamma$. The peaks initially move away from each other, one more into the gap and the other farther away from it as the system gradually thermalizes.  At time $t\approx500$, the trend at $\Gamma$ reverses and the main QP peaks start to move back towards their equilibrium position. These dynamics are however distinct for the $X$ point, for which the splitting keeps growing and only at the end of the simulation time, the QP peaks positions stabilize. 

Overall, our calculations clearly demonstrate that the evolution of the electronic structure is nontrivially related to the non-equilibrium time-dependent distribution of charge carriers and the induced dynamical correlations. The presence of long-range density-density Coulomb coupling among the photoexcited electrons in the conduction states leads to the emergence of (plasmon-like) satellites. Upon thermalization (i.e., condensation of electrons at the band edge), the satellite strength gradually decreases. This is further supported by the lack of strong satellites states for $U_{\rm intra} =1$.


\section{Conclusion and Outlook}
In this work, we have presented a new development based on two recently introduced approaches: RT-DE allowing an efficient description of non-equilibrium spectral functions including (induced) electronic correlations combined with the Lindblad-Kadanoff-Baym approach providing a new formalism for the diagrammatic treatment of open quantum systems dynamics.  Specifically, we have developed Lindblad MBPT for two-body particle-hole loss operators representing interactions with a bosonic bath and included their contributions to the system's self-energy.  In practice, the coupling represents energy transfer to a phonon bath leading to a gradual condensation of charge carriers at the band edges. This is further combined with a photon bath allowing for radiative recombination.

We have applied this methodology to study the time-dependent renormalization of excited-state spectra for a model semiconductor system which includes long-range Coulombic interactions. The system is pulsed with light driving the transition of a realistic fraction of electrons ($\sim2\%$) between the valence and conduction states at finite $k$.  Our results show first an abrupt change in the bandstructure followed by a dynamical, monotonic shrinking of the bandgap as electrons thermalize.  We observe that this behavior is strongly $k$-dependent, with the bands at certain points in the Brillouin zone moving farther away from the chemical potential, and some moving inwards illustrating the nontrivial nature of the couplings between dissipation and electronic corralation effects.  We also see $k$-dependent branching of QP peaks analogous to non-equilibrium and interaction driven gap opening. This is accompanied by the formation of short lived satellite peaks in the excited state spectra. The presented computational framework enables to capture the evolution of QP state energy shifts, as well as state emergence and dissapearance depending on the non-equilibrium electron distribution.  

The observed dynamics is nontrivial despite we utilized the simplest form of the Lindblad operators, which can be further extended to describe a large variety of external baths.  Such extension will make this methodology important for analyzing many questions in non-equilibrium open quantum systems which have been out of reach for the NEGF formalism for extended systems up to now. This includes phenomena such as the thermalization of excitonic states, electron-phonon and electron-magnon coupling in realistic materials, and spin dynamics influenced by chiral baths in the chirality-induced spin selectivity effects.

In situations where the dissipative strength become similar to the Coulombic interactions, the self-energy approximations must be extended to include diagrams which are higher order in the Lindblad couplings. The higher order dissipative self-energy diagrams, beyond the lowest order considered here, include additional dynamical effects, e.g., as the direct feedback with the bath, and they will be presented in future work. The key advantage of the presented framework is that RTDE readily enables to incorporate the extended Lindblad-Kadanoff-Baym formulation and treat electron-electron and electron-bath scatterings on equal footing.

\section{Acknowledgments}
This material is based upon work supported by the U.S. Department of Energy, Office of Science, Office of Advanced Scientific Computing Research and Office of Basic Energy Sciences, Scientific Discovery through Advanced Computing (SciDAC) program under Award Number DE-SC0022198. This research used resources of the National Energy Research Scientific Computing Center, a DOE Office of Science User Facility supported by the Office of Science of the U.S. Department of Energy under Contract No. DE-AC02-05CH11231 using NERSC Award No. BES-ERCAP0032056.

GS and EP acknowledge funding from Ministero Università e
Ricerca PRIN under Grant Agreement No. 2022WZ8LME,
from INFN through project TIME2QUEST, from European Research Council MSCA-ITN TIMES under Grant Agreement No. 101118915, and from Tor Vergata University through project TESLA.

\bibliography{mybib}

\begin{thebibliography}{48}%
\makeatletter
\providecommand \@ifxundefined [1]{%
 \@ifx{#1\undefined}
}%
\providecommand \@ifnum [1]{%
 \ifnum #1\expandafter \@firstoftwo
 \else \expandafter \@secondoftwo
 \fi
}%
\providecommand \@ifx [1]{%
 \ifx #1\expandafter \@firstoftwo
 \else \expandafter \@secondoftwo
 \fi
}%
\providecommand \natexlab [1]{#1}%
\providecommand \enquote  [1]{``#1''}%
\providecommand \bibnamefont  [1]{#1}%
\providecommand \bibfnamefont [1]{#1}%
\providecommand \citenamefont [1]{#1}%
\providecommand \href@noop [0]{\@secondoftwo}%
\providecommand \href [0]{\begingroup \@sanitize@url \@href}%
\providecommand \@href[1]{\@@startlink{#1}\@@href}%
\providecommand \@@href[1]{\endgroup#1\@@endlink}%
\providecommand \@sanitize@url [0]{\catcode `\\12\catcode `\$12\catcode
  `\&12\catcode `\#12\catcode `\^12\catcode `\_12\catcode `\%12\relax}%
\providecommand \@@startlink[1]{}%
\providecommand \@@endlink[0]{}%
\providecommand \url  [0]{\begingroup\@sanitize@url \@url }%
\providecommand \@url [1]{\endgroup\@href {#1}{\urlprefix }}%
\providecommand \urlprefix  [0]{URL }%
\providecommand \Eprint [0]{\href }%
\providecommand \doibase [0]{https://doi.org/}%
\providecommand \selectlanguage [0]{\@gobble}%
\providecommand \bibinfo  [0]{\@secondoftwo}%
\providecommand \bibfield  [0]{\@secondoftwo}%
\providecommand \translation [1]{[#1]}%
\providecommand \BibitemOpen [0]{}%
\providecommand \bibitemStop [0]{}%
\providecommand \bibitemNoStop [0]{.\EOS\space}%
\providecommand \EOS [0]{\spacefactor3000\relax}%
\providecommand \BibitemShut  [1]{\csname bibitem#1\endcsname}%
\let\auto@bib@innerbib\@empty
\bibitem [{\citenamefont {Baumann}\ \emph {et~al.}(2010)\citenamefont
  {Baumann}, \citenamefont {Guerlin}, \citenamefont {Brennecke},\ and\
  \citenamefont {Esslinger}}]{Baumann2010}%
  \BibitemOpen
  \bibfield  {author} {\bibinfo {author} {\bibfnamefont {K.}~\bibnamefont
  {Baumann}}, \bibinfo {author} {\bibfnamefont {C.}~\bibnamefont {Guerlin}},
  \bibinfo {author} {\bibfnamefont {F.}~\bibnamefont {Brennecke}},\ and\
  \bibinfo {author} {\bibfnamefont {T.}~\bibnamefont {Esslinger}},\ }\href
  {https://doi.org/10.1038/nature09009} {\bibfield  {journal} {\bibinfo
  {journal} {Nature}\ }\textbf {\bibinfo {volume} {464}},\ \bibinfo {pages}
  {1301} (\bibinfo {year} {2010})}\BibitemShut {NoStop}%
\bibitem [{\citenamefont {Ritsch}\ \emph {et~al.}(2013)\citenamefont {Ritsch},
  \citenamefont {Domokos}, \citenamefont {Brennecke},\ and\ \citenamefont
  {Esslinger}}]{Ritsch2013}%
  \BibitemOpen
  \bibfield  {author} {\bibinfo {author} {\bibfnamefont {H.}~\bibnamefont
  {Ritsch}}, \bibinfo {author} {\bibfnamefont {P.}~\bibnamefont {Domokos}},
  \bibinfo {author} {\bibfnamefont {F.}~\bibnamefont {Brennecke}},\ and\
  \bibinfo {author} {\bibfnamefont {T.}~\bibnamefont {Esslinger}},\ }\href
  {https://doi.org/10.1103/RevModPhys.85.553} {\bibfield  {journal} {\bibinfo
  {journal} {Rev. Mod. Phys.}\ }\textbf {\bibinfo {volume} {85}},\ \bibinfo
  {pages} {553} (\bibinfo {year} {2013})}\BibitemShut {NoStop}%
\bibitem [{\citenamefont {Mori}(2023)}]{Mori2023}%
  \BibitemOpen
  \bibfield  {author} {\bibinfo {author} {\bibfnamefont {T.}~\bibnamefont
  {Mori}},\ }\href
  {https://doi.org/https://doi.org/10.1146/annurev-conmatphys-040721-015537}
  {\bibfield  {journal} {\bibinfo  {journal} {Annual Review of Condensed Matter
  Physics}\ }\textbf {\bibinfo {volume} {14}},\ \bibinfo {pages} {35} (\bibinfo
  {year} {2023})}\BibitemShut {NoStop}%
\bibitem [{\citenamefont {Sato}\ \emph {et~al.}(2020)\citenamefont {Sato},
  \citenamefont {De~Giovannini}, \citenamefont {Aeschlimann}, \citenamefont
  {Gierz}, \citenamefont {Hübener},\ and\ \citenamefont {Rubio}}]{Sato_2020}%
  \BibitemOpen
  \bibfield  {author} {\bibinfo {author} {\bibfnamefont {S.~A.}\ \bibnamefont
  {Sato}}, \bibinfo {author} {\bibfnamefont {U.}~\bibnamefont {De~Giovannini}},
  \bibinfo {author} {\bibfnamefont {S.}~\bibnamefont {Aeschlimann}}, \bibinfo
  {author} {\bibfnamefont {I.}~\bibnamefont {Gierz}}, \bibinfo {author}
  {\bibfnamefont {H.}~\bibnamefont {Hübener}},\ and\ \bibinfo {author}
  {\bibfnamefont {A.}~\bibnamefont {Rubio}},\ }\href
  {https://doi.org/10.1088/1361-6455/abb127} {\bibfield  {journal} {\bibinfo
  {journal} {Journal of Physics B: Atomic, Molecular and Optical Physics}\
  }\textbf {\bibinfo {volume} {53}},\ \bibinfo {pages} {225601} (\bibinfo
  {year} {2020})}\BibitemShut {NoStop}%
\bibitem [{\citenamefont {Wouters}\ and\ \citenamefont
  {Carusotto}(2007)}]{Wouters2007}%
  \BibitemOpen
  \bibfield  {author} {\bibinfo {author} {\bibfnamefont {M.}~\bibnamefont
  {Wouters}}\ and\ \bibinfo {author} {\bibfnamefont {I.}~\bibnamefont
  {Carusotto}},\ }\href {https://doi.org/10.1103/PhysRevLett.99.140402}
  {\bibfield  {journal} {\bibinfo  {journal} {Phys. Rev. Lett.}\ }\textbf
  {\bibinfo {volume} {99}},\ \bibinfo {pages} {140402} (\bibinfo {year}
  {2007})}\BibitemShut {NoStop}%
\bibitem [{\citenamefont {Deng}\ \emph {et~al.}(2010)\citenamefont {Deng},
  \citenamefont {Haug},\ and\ \citenamefont {Yamamoto}}]{Deng2010}%
  \BibitemOpen
  \bibfield  {author} {\bibinfo {author} {\bibfnamefont {H.}~\bibnamefont
  {Deng}}, \bibinfo {author} {\bibfnamefont {H.}~\bibnamefont {Haug}},\ and\
  \bibinfo {author} {\bibfnamefont {Y.}~\bibnamefont {Yamamoto}},\ }\href
  {https://doi.org/10.1103/RevModPhys.82.1489} {\bibfield  {journal} {\bibinfo
  {journal} {Rev. Mod. Phys.}\ }\textbf {\bibinfo {volume} {82}},\ \bibinfo
  {pages} {1489} (\bibinfo {year} {2010})}\BibitemShut {NoStop}%
\bibitem [{\citenamefont {Byrnes}\ \emph {et~al.}(2014)\citenamefont {Byrnes},
  \citenamefont {Kim},\ and\ \citenamefont {Yamamoto}}]{Byrnes2014}%
  \BibitemOpen
  \bibfield  {author} {\bibinfo {author} {\bibfnamefont {T.}~\bibnamefont
  {Byrnes}}, \bibinfo {author} {\bibfnamefont {N.~Y.}\ \bibnamefont {Kim}},\
  and\ \bibinfo {author} {\bibfnamefont {Y.}~\bibnamefont {Yamamoto}},\ }\href
  {https://doi.org/10.1038/nphys3143} {\bibfield  {journal} {\bibinfo
  {journal} {Nature Physics}\ }\textbf {\bibinfo {volume} {10}},\ \bibinfo
  {pages} {803} (\bibinfo {year} {2014})}\BibitemShut {NoStop}%
\bibitem [{\citenamefont {Kasprzak}\ \emph {et~al.}(2006)\citenamefont
  {Kasprzak}, \citenamefont {Richard}, \citenamefont {Kundermann},
  \citenamefont {Baas}, \citenamefont {Jeambrun}, \citenamefont {Keeling},
  \citenamefont {Marchetti}, \citenamefont {Szyma{\'{n}}ska}, \citenamefont
  {Andr{\'e}}, \citenamefont {Staehli}, \citenamefont {Savona}, \citenamefont
  {Littlewood}, \citenamefont {Deveaud},\ and\ \citenamefont
  {Dang}}]{Kasprzak2006}%
  \BibitemOpen
  \bibfield  {author} {\bibinfo {author} {\bibfnamefont {J.}~\bibnamefont
  {Kasprzak}}, \bibinfo {author} {\bibfnamefont {M.}~\bibnamefont {Richard}},
  \bibinfo {author} {\bibfnamefont {S.}~\bibnamefont {Kundermann}}, \bibinfo
  {author} {\bibfnamefont {A.}~\bibnamefont {Baas}}, \bibinfo {author}
  {\bibfnamefont {P.}~\bibnamefont {Jeambrun}}, \bibinfo {author}
  {\bibfnamefont {J.~M.~J.}\ \bibnamefont {Keeling}}, \bibinfo {author}
  {\bibfnamefont {F.~M.}\ \bibnamefont {Marchetti}}, \bibinfo {author}
  {\bibfnamefont {M.~H.}\ \bibnamefont {Szyma{\'{n}}ska}}, \bibinfo {author}
  {\bibfnamefont {R.}~\bibnamefont {Andr{\'e}}}, \bibinfo {author}
  {\bibfnamefont {J.~L.}\ \bibnamefont {Staehli}}, \bibinfo {author}
  {\bibfnamefont {V.}~\bibnamefont {Savona}}, \bibinfo {author} {\bibfnamefont
  {P.~B.}\ \bibnamefont {Littlewood}}, \bibinfo {author} {\bibfnamefont
  {B.}~\bibnamefont {Deveaud}},\ and\ \bibinfo {author} {\bibfnamefont {L.~S.}\
  \bibnamefont {Dang}},\ }\href {https://doi.org/10.1038/nature05131}
  {\bibfield  {journal} {\bibinfo  {journal} {Nature}\ }\textbf {\bibinfo
  {volume} {443}},\ \bibinfo {pages} {409} (\bibinfo {year}
  {2006})}\BibitemShut {NoStop}%
\bibitem [{\citenamefont {Krauter}\ \emph {et~al.}(2011)\citenamefont
  {Krauter}, \citenamefont {Muschik}, \citenamefont {Jensen}, \citenamefont
  {Wasilewski}, \citenamefont {Petersen}, \citenamefont {Cirac},\ and\
  \citenamefont {Polzik}}]{Krauter2011}%
  \BibitemOpen
  \bibfield  {author} {\bibinfo {author} {\bibfnamefont {H.}~\bibnamefont
  {Krauter}}, \bibinfo {author} {\bibfnamefont {C.~A.}\ \bibnamefont
  {Muschik}}, \bibinfo {author} {\bibfnamefont {K.}~\bibnamefont {Jensen}},
  \bibinfo {author} {\bibfnamefont {W.}~\bibnamefont {Wasilewski}}, \bibinfo
  {author} {\bibfnamefont {J.~M.}\ \bibnamefont {Petersen}}, \bibinfo {author}
  {\bibfnamefont {J.~I.}\ \bibnamefont {Cirac}},\ and\ \bibinfo {author}
  {\bibfnamefont {E.~S.}\ \bibnamefont {Polzik}},\ }\href
  {https://doi.org/10.1103/PhysRevLett.107.080503} {\bibfield  {journal}
  {\bibinfo  {journal} {Phys. Rev. Lett.}\ }\textbf {\bibinfo {volume} {107}},\
  \bibinfo {pages} {080503} (\bibinfo {year} {2011})}\BibitemShut {NoStop}%
\bibitem [{\citenamefont {Kastoryano}\ \emph {et~al.}(2011)\citenamefont
  {Kastoryano}, \citenamefont {Reiter},\ and\ \citenamefont
  {S\o{}rensen}}]{Kastoryano2011}%
  \BibitemOpen
  \bibfield  {author} {\bibinfo {author} {\bibfnamefont {M.~J.}\ \bibnamefont
  {Kastoryano}}, \bibinfo {author} {\bibfnamefont {F.}~\bibnamefont {Reiter}},\
  and\ \bibinfo {author} {\bibfnamefont {A.~S.}\ \bibnamefont {S\o{}rensen}},\
  }\href {https://doi.org/10.1103/PhysRevLett.106.090502} {\bibfield  {journal}
  {\bibinfo  {journal} {Phys. Rev. Lett.}\ }\textbf {\bibinfo {volume} {106}},\
  \bibinfo {pages} {090502} (\bibinfo {year} {2011})}\BibitemShut {NoStop}%
\bibitem [{\citenamefont {Del~Re}\ \emph {et~al.}(2020)\citenamefont {Del~Re},
  \citenamefont {Rost}, \citenamefont {Kemper},\ and\ \citenamefont
  {Freericks}}]{DelRe2020}%
  \BibitemOpen
  \bibfield  {author} {\bibinfo {author} {\bibfnamefont {L.}~\bibnamefont
  {Del~Re}}, \bibinfo {author} {\bibfnamefont {B.}~\bibnamefont {Rost}},
  \bibinfo {author} {\bibfnamefont {A.~F.}\ \bibnamefont {Kemper}},\ and\
  \bibinfo {author} {\bibfnamefont {J.~K.}\ \bibnamefont {Freericks}},\ }\href
  {https://doi.org/10.1103/PhysRevB.102.125112} {\bibfield  {journal} {\bibinfo
   {journal} {Phys. Rev. B}\ }\textbf {\bibinfo {volume} {102}},\ \bibinfo
  {pages} {125112} (\bibinfo {year} {2020})}\BibitemShut {NoStop}%
\bibitem [{\citenamefont {Del~Re}\ \emph {et~al.}(2024)\citenamefont {Del~Re},
  \citenamefont {Rost}, \citenamefont {Foss-Feig}, \citenamefont {Kemper},\
  and\ \citenamefont {Freericks}}]{DelRe2024}%
  \BibitemOpen
  \bibfield  {author} {\bibinfo {author} {\bibfnamefont {L.}~\bibnamefont
  {Del~Re}}, \bibinfo {author} {\bibfnamefont {B.}~\bibnamefont {Rost}},
  \bibinfo {author} {\bibfnamefont {M.}~\bibnamefont {Foss-Feig}}, \bibinfo
  {author} {\bibfnamefont {A.~F.}\ \bibnamefont {Kemper}},\ and\ \bibinfo
  {author} {\bibfnamefont {J.~K.}\ \bibnamefont {Freericks}},\ }\href
  {https://doi.org/10.1103/PhysRevLett.132.100601} {\bibfield  {journal}
  {\bibinfo  {journal} {Phys. Rev. Lett.}\ }\textbf {\bibinfo {volume} {132}},\
  \bibinfo {pages} {100601} (\bibinfo {year} {2024})}\BibitemShut {NoStop}%
\bibitem [{\citenamefont {Rost}\ \emph {et~al.}(2025)\citenamefont {Rost},
  \citenamefont {Del~Re}, \citenamefont {Earnest}, \citenamefont {Kemper},
  \citenamefont {Jones},\ and\ \citenamefont {Freericks}}]{Rost2025}%
  \BibitemOpen
  \bibfield  {author} {\bibinfo {author} {\bibfnamefont {B.}~\bibnamefont
  {Rost}}, \bibinfo {author} {\bibfnamefont {L.}~\bibnamefont {Del~Re}},
  \bibinfo {author} {\bibfnamefont {N.}~\bibnamefont {Earnest}}, \bibinfo
  {author} {\bibfnamefont {A.~F.}\ \bibnamefont {Kemper}}, \bibinfo {author}
  {\bibfnamefont {B.}~\bibnamefont {Jones}},\ and\ \bibinfo {author}
  {\bibfnamefont {J.~K.}\ \bibnamefont {Freericks}},\ }\href
  {https://doi.org/10.1038/s41534-025-00964-8} {\bibfield  {journal} {\bibinfo
  {journal} {npj Quantum Information}\ }\textbf {\bibinfo {volume} {11}},\
  \bibinfo {pages} {10} (\bibinfo {year} {2025})}\BibitemShut {NoStop}%
\bibitem [{\citenamefont {Guo}\ \emph {et~al.}(2025)\citenamefont {Guo},
  \citenamefont {Riva}, \citenamefont {Simoni}, \citenamefont {Xu},\ and\
  \citenamefont {Ping}}]{guo2025phononassistedradiativelifetimesexciton}%
  \BibitemOpen
  \bibfield  {author} {\bibinfo {author} {\bibfnamefont {C.}~\bibnamefont
  {Guo}}, \bibinfo {author} {\bibfnamefont {G.}~\bibnamefont {Riva}}, \bibinfo
  {author} {\bibfnamefont {J.}~\bibnamefont {Simoni}}, \bibinfo {author}
  {\bibfnamefont {J.}~\bibnamefont {Xu}},\ and\ \bibinfo {author}
  {\bibfnamefont {Y.}~\bibnamefont {Ping}},\ }\href
  {https://arxiv.org/abs/2504.18071} {\bibinfo {title} {Phonon-assisted
  radiative lifetimes and exciton dynamics from first principles}} (\bibinfo
  {year} {2025}),\ \Eprint {https://arxiv.org/abs/2504.18071} {arXiv:2504.18071
  [cond-mat.mtrl-sci]} \BibitemShut {NoStop}%
\bibitem [{\citenamefont {Simoni}\ \emph {et~al.}(2025)\citenamefont {Simoni},
  \citenamefont {Riva},\ and\ \citenamefont
  {Ping}}]{simoni2025firstprinciplesopenquantumdynamics}%
  \BibitemOpen
  \bibfield  {author} {\bibinfo {author} {\bibfnamefont {J.}~\bibnamefont
  {Simoni}}, \bibinfo {author} {\bibfnamefont {G.}~\bibnamefont {Riva}},\ and\
  \bibinfo {author} {\bibfnamefont {Y.}~\bibnamefont {Ping}},\ }\href
  {https://arxiv.org/abs/2504.17936} {\bibinfo {title} {First-principles open
  quantum dynamics for solids based on density-matrix formalism}} (\bibinfo
  {year} {2025}),\ \Eprint {https://arxiv.org/abs/2504.17936} {arXiv:2504.17936
  [cond-mat.mtrl-sci]} \BibitemShut {NoStop}%
\bibitem [{\citenamefont {Reeves}\ \emph {et~al.}(2025)\citenamefont {Reeves},
  \citenamefont {Cushing},\ and\ \citenamefont {Vlcek}}]{reeves2025}%
  \BibitemOpen
  \bibfield  {author} {\bibinfo {author} {\bibfnamefont {C.~C.}\ \bibnamefont
  {Reeves}}, \bibinfo {author} {\bibfnamefont {S.~K.}\ \bibnamefont
  {Cushing}},\ and\ \bibinfo {author} {\bibfnamefont {V.}~\bibnamefont
  {Vlcek}},\ }\href {https://arxiv.org/abs/2503.09715} {\bibinfo {title} {The
  role of effective mass and long-range interactions in the band-gap
  renormalization of photo-excited semiconductors}} (\bibinfo {year} {2025}),\
  \Eprint {https://arxiv.org/abs/2503.09715} {arXiv:2503.09715
  [cond-mat.mtrl-sci]} \BibitemShut {NoStop}%
\bibitem [{\citenamefont {Chernikov}\ \emph {et~al.}(2015)\citenamefont
  {Chernikov}, \citenamefont {Ruppert}, \citenamefont {Hill}, \citenamefont
  {Rigosi},\ and\ \citenamefont {Heinz}}]{Chernikov_2015_2}%
  \BibitemOpen
  \bibfield  {author} {\bibinfo {author} {\bibfnamefont {A.}~\bibnamefont
  {Chernikov}}, \bibinfo {author} {\bibfnamefont {C.}~\bibnamefont {Ruppert}},
  \bibinfo {author} {\bibfnamefont {H.~M.}\ \bibnamefont {Hill}}, \bibinfo
  {author} {\bibfnamefont {A.~F.}\ \bibnamefont {Rigosi}},\ and\ \bibinfo
  {author} {\bibfnamefont {T.~F.}\ \bibnamefont {Heinz}},\ }\href
  {https://doi.org/10.1038/nphoton.2015.104} {\bibfield  {journal} {\bibinfo
  {journal} {Nature Photonics}\ }\textbf {\bibinfo {volume} {9}},\ \bibinfo
  {pages} {466} (\bibinfo {year} {2015})}\BibitemShut {NoStop}%
\bibitem [{\citenamefont {Ulstrup}\ \emph {et~al.}(2016)\citenamefont
  {Ulstrup}, \citenamefont {Čabo}, \citenamefont {Miwa}, \citenamefont
  {Riley}, \citenamefont {Grønborg}, \citenamefont {Johannsen}, \citenamefont
  {Cacho}, \citenamefont {Alexander}, \citenamefont {Chapman}, \citenamefont
  {Springate}, \citenamefont {Bianchi}, \citenamefont {Dendzik}, \citenamefont
  {Lauritsen}, \citenamefont {King},\ and\ \citenamefont
  {Hofmann}}]{Ulstrop_2016}%
  \BibitemOpen
  \bibfield  {author} {\bibinfo {author} {\bibfnamefont {S.}~\bibnamefont
  {Ulstrup}}, \bibinfo {author} {\bibfnamefont {A.~G.}\ \bibnamefont {Čabo}},
  \bibinfo {author} {\bibfnamefont {J.~A.}\ \bibnamefont {Miwa}}, \bibinfo
  {author} {\bibfnamefont {J.~M.}\ \bibnamefont {Riley}}, \bibinfo {author}
  {\bibfnamefont {S.~S.}\ \bibnamefont {Grønborg}}, \bibinfo {author}
  {\bibfnamefont {J.~C.}\ \bibnamefont {Johannsen}}, \bibinfo {author}
  {\bibfnamefont {C.}~\bibnamefont {Cacho}}, \bibinfo {author} {\bibfnamefont
  {O.}~\bibnamefont {Alexander}}, \bibinfo {author} {\bibfnamefont {R.~T.}\
  \bibnamefont {Chapman}}, \bibinfo {author} {\bibfnamefont {E.}~\bibnamefont
  {Springate}}, \bibinfo {author} {\bibfnamefont {M.}~\bibnamefont {Bianchi}},
  \bibinfo {author} {\bibfnamefont {M.}~\bibnamefont {Dendzik}}, \bibinfo
  {author} {\bibfnamefont {J.~V.}\ \bibnamefont {Lauritsen}}, \bibinfo {author}
  {\bibfnamefont {P.~D.~C.}\ \bibnamefont {King}},\ and\ \bibinfo {author}
  {\bibfnamefont {P.}~\bibnamefont {Hofmann}},\ }\href
  {https://doi.org/10.1021/acsnano.6b02622} {\bibfield  {journal} {\bibinfo
  {journal} {ACS Nano}\ }\textbf {\bibinfo {volume} {10}},\ \bibinfo {pages}
  {6315} (\bibinfo {year} {2016})}\BibitemShut {NoStop}%
\bibitem [{\citenamefont {{Roth}}\ \emph {et~al.}(2019)\citenamefont {{Roth}},
  \citenamefont {{Crepaldi}}, \citenamefont {{Puppin}}, \citenamefont
  {{Gatti}}, \citenamefont {{Bugini}}, \citenamefont {{Grimaldi}},
  \citenamefont {{Barrilot}}, \citenamefont {{Arrell}}, \citenamefont
  {{Frassetto}}, \citenamefont {{Poletto}}, \citenamefont {{Chergui}},
  \citenamefont {{Marini}},\ and\ \citenamefont {{Grioni}}}]{Roth_2019}%
  \BibitemOpen
  \bibfield  {author} {\bibinfo {author} {\bibfnamefont {S.}~\bibnamefont
  {{Roth}}}, \bibinfo {author} {\bibfnamefont {A.}~\bibnamefont {{Crepaldi}}},
  \bibinfo {author} {\bibfnamefont {M.}~\bibnamefont {{Puppin}}}, \bibinfo
  {author} {\bibfnamefont {G.}~\bibnamefont {{Gatti}}}, \bibinfo {author}
  {\bibfnamefont {D.}~\bibnamefont {{Bugini}}}, \bibinfo {author}
  {\bibfnamefont {I.}~\bibnamefont {{Grimaldi}}}, \bibinfo {author}
  {\bibfnamefont {T.~R.}\ \bibnamefont {{Barrilot}}}, \bibinfo {author}
  {\bibfnamefont {C.~A.}\ \bibnamefont {{Arrell}}}, \bibinfo {author}
  {\bibfnamefont {F.}~\bibnamefont {{Frassetto}}}, \bibinfo {author}
  {\bibfnamefont {L.}~\bibnamefont {{Poletto}}}, \bibinfo {author}
  {\bibfnamefont {M.}~\bibnamefont {{Chergui}}}, \bibinfo {author}
  {\bibfnamefont {A.}~\bibnamefont {{Marini}}},\ and\ \bibinfo {author}
  {\bibfnamefont {M.}~\bibnamefont {{Grioni}}},\ }\href
  {https://doi.org/10.1088/2053-1583/ab1216} {\bibfield  {journal} {\bibinfo
  {journal} {2D Materials}\ }\textbf {\bibinfo {volume} {6}},\ \bibinfo {eid}
  {031001} (\bibinfo {year} {2019})}\BibitemShut {NoStop}%
\bibitem [{\citenamefont {Kang}\ \emph {et~al.}(2017)\citenamefont {Kang},
  \citenamefont {Kim}, \citenamefont {Ryu}, \citenamefont {Jung}, \citenamefont
  {Kim}, \citenamefont {Moreschini}, \citenamefont {Jozwiak}, \citenamefont
  {Rotenberg}, \citenamefont {Bostwick},\ and\ \citenamefont
  {Kim}}]{Kang_2017}%
  \BibitemOpen
  \bibfield  {author} {\bibinfo {author} {\bibfnamefont {M.}~\bibnamefont
  {Kang}}, \bibinfo {author} {\bibfnamefont {B.}~\bibnamefont {Kim}}, \bibinfo
  {author} {\bibfnamefont {S.~H.}\ \bibnamefont {Ryu}}, \bibinfo {author}
  {\bibfnamefont {S.~W.}\ \bibnamefont {Jung}}, \bibinfo {author}
  {\bibfnamefont {J.}~\bibnamefont {Kim}}, \bibinfo {author} {\bibfnamefont
  {L.}~\bibnamefont {Moreschini}}, \bibinfo {author} {\bibfnamefont
  {C.}~\bibnamefont {Jozwiak}}, \bibinfo {author} {\bibfnamefont
  {E.}~\bibnamefont {Rotenberg}}, \bibinfo {author} {\bibfnamefont
  {A.}~\bibnamefont {Bostwick}},\ and\ \bibinfo {author} {\bibfnamefont
  {K.~S.}\ \bibnamefont {Kim}},\ }\href
  {https://doi.org/10.1021/acs.nanolett.6b04775} {\bibfield  {journal}
  {\bibinfo  {journal} {Nano Letters}\ }\textbf {\bibinfo {volume} {17}},\
  \bibinfo {pages} {1610} (\bibinfo {year} {2017})}\BibitemShut {NoStop}%
\bibitem [{\citenamefont {Boschini}\ \emph {et~al.}(2024)\citenamefont
  {Boschini}, \citenamefont {Zonno},\ and\ \citenamefont
  {Damascelli}}]{Boschini2024}%
  \BibitemOpen
  \bibfield  {author} {\bibinfo {author} {\bibfnamefont {F.}~\bibnamefont
  {Boschini}}, \bibinfo {author} {\bibfnamefont {M.}~\bibnamefont {Zonno}},\
  and\ \bibinfo {author} {\bibfnamefont {A.}~\bibnamefont {Damascelli}},\
  }\href {https://doi.org/10.1103/RevModPhys.96.015003} {\bibfield  {journal}
  {\bibinfo  {journal} {Rev. Mod. Phys.}\ }\textbf {\bibinfo {volume} {96}},\
  \bibinfo {pages} {015003} (\bibinfo {year} {2024})}\BibitemShut {NoStop}%
\bibitem [{\citenamefont {Stefanucci}\ and\ \citenamefont {van
  Leeuwen}(2013)}]{SvL2013}%
  \BibitemOpen
  \bibfield  {author} {\bibinfo {author} {\bibfnamefont {G.}~\bibnamefont
  {Stefanucci}}\ and\ \bibinfo {author} {\bibfnamefont {R.}~\bibnamefont {van
  Leeuwen}},\ }\href@noop {} {\emph {\bibinfo {title} {Nonequilibrium Many-Body
  Theory of Quantum Systems}}}\ (\bibinfo  {publisher} {Cambridge University
  Press},\ \bibinfo {address} {Cambridge, UK},\ \bibinfo {year}
  {2013})\BibitemShut {NoStop}%
\bibitem [{\citenamefont {Sieberer}\ \emph {et~al.}(2016)\citenamefont
  {Sieberer}, \citenamefont {Buchhold},\ and\ \citenamefont
  {Diehl}}]{Sieberer_2016}%
  \BibitemOpen
  \bibfield  {author} {\bibinfo {author} {\bibfnamefont {L.~M.}\ \bibnamefont
  {Sieberer}}, \bibinfo {author} {\bibfnamefont {M.}~\bibnamefont {Buchhold}},\
  and\ \bibinfo {author} {\bibfnamefont {S.}~\bibnamefont {Diehl}},\ }\href
  {https://doi.org/10.1088/0034-4885/79/9/096001} {\bibfield  {journal}
  {\bibinfo  {journal} {Reports on Progress in Physics}\ }\textbf {\bibinfo
  {volume} {79}},\ \bibinfo {pages} {096001} (\bibinfo {year}
  {2016})}\BibitemShut {NoStop}%
\bibitem [{\citenamefont {Sieberer}\ \emph {et~al.}(2023)\citenamefont
  {Sieberer}, \citenamefont {Buchhold}, \citenamefont {Marino},\ and\
  \citenamefont {Diehl}}]{sieberer2023}%
  \BibitemOpen
  \bibfield  {author} {\bibinfo {author} {\bibfnamefont {L.~M.}\ \bibnamefont
  {Sieberer}}, \bibinfo {author} {\bibfnamefont {M.}~\bibnamefont {Buchhold}},
  \bibinfo {author} {\bibfnamefont {J.}~\bibnamefont {Marino}},\ and\ \bibinfo
  {author} {\bibfnamefont {S.}~\bibnamefont {Diehl}},\ }\href
  {https://arxiv.org/abs/2312.03073} {\bibinfo {title} {Universality in driven
  open quantum matter}} (\bibinfo {year} {2023}),\ \Eprint
  {https://arxiv.org/abs/2312.03073} {arXiv:2312.03073 [cond-mat.stat-mech]}
  \BibitemShut {NoStop}%
\bibitem [{\citenamefont {Fogedby}(2022)}]{Fogedby2022}%
  \BibitemOpen
  \bibfield  {author} {\bibinfo {author} {\bibfnamefont {H.~C.}\ \bibnamefont
  {Fogedby}},\ }\href {https://doi.org/10.1103/PhysRevA.106.022205} {\bibfield
  {journal} {\bibinfo  {journal} {Phys. Rev. A}\ }\textbf {\bibinfo {volume}
  {106}},\ \bibinfo {pages} {022205} (\bibinfo {year} {2022})}\BibitemShut
  {NoStop}%
\bibitem [{\citenamefont {Thompson}\ and\ \citenamefont
  {Kamenev}(2023)}]{Thompson2023}%
  \BibitemOpen
  \bibfield  {author} {\bibinfo {author} {\bibfnamefont {F.}~\bibnamefont
  {Thompson}}\ and\ \bibinfo {author} {\bibfnamefont {A.}~\bibnamefont
  {Kamenev}},\ }\href
  {https://doi.org/https://doi.org/10.1016/j.aop.2023.169385} {\bibfield
  {journal} {\bibinfo  {journal} {Annals of Physics}\ }\textbf {\bibinfo
  {volume} {455}},\ \bibinfo {pages} {169385} (\bibinfo {year}
  {2023})}\BibitemShut {NoStop}%
\bibitem [{\citenamefont {Stefanucci}(2024)}]{Stefanucci2024}%
  \BibitemOpen
  \bibfield  {author} {\bibinfo {author} {\bibfnamefont {G.}~\bibnamefont
  {Stefanucci}},\ }\href {https://doi.org/10.1103/PhysRevLett.133.066901}
  {\bibfield  {journal} {\bibinfo  {journal} {Phys. Rev. Lett.}\ }\textbf
  {\bibinfo {volume} {133}},\ \bibinfo {pages} {066901} (\bibinfo {year}
  {2024})}\BibitemShut {NoStop}%
\bibitem [{\citenamefont {Sch\"uler}\ \emph {et~al.}(2018)\citenamefont
  {Sch\"uler}, \citenamefont {Eckstein},\ and\ \citenamefont
  {Werner}}]{Truncation}%
  \BibitemOpen
  \bibfield  {author} {\bibinfo {author} {\bibfnamefont {M.}~\bibnamefont
  {Sch\"uler}}, \bibinfo {author} {\bibfnamefont {M.}~\bibnamefont
  {Eckstein}},\ and\ \bibinfo {author} {\bibfnamefont {P.}~\bibnamefont
  {Werner}},\ }\href {https://doi.org/10.1103/PhysRevB.97.245129} {\bibfield
  {journal} {\bibinfo  {journal} {Phys. Rev. B}\ }\textbf {\bibinfo {volume}
  {97}},\ \bibinfo {pages} {245129} (\bibinfo {year} {2018})}\BibitemShut
  {NoStop}%
\bibitem [{\citenamefont {Picano}\ and\ \citenamefont
  {Eckstein}(2021)}]{AcceleratedCollapse}%
  \BibitemOpen
  \bibfield  {author} {\bibinfo {author} {\bibfnamefont {A.}~\bibnamefont
  {Picano}}\ and\ \bibinfo {author} {\bibfnamefont {M.}~\bibnamefont
  {Eckstein}},\ }\href {https://doi.org/10.1103/PhysRevB.103.165118} {\bibfield
   {journal} {\bibinfo  {journal} {Phys. Rev. B}\ }\textbf {\bibinfo {volume}
  {103}},\ \bibinfo {pages} {165118} (\bibinfo {year} {2021})}\BibitemShut
  {NoStop}%
\bibitem [{\citenamefont {Kaye}\ and\ \citenamefont
  {Golež}(2021)}]{KayeComp2021}%
  \BibitemOpen
  \bibfield  {author} {\bibinfo {author} {\bibfnamefont {J.}~\bibnamefont
  {Kaye}}\ and\ \bibinfo {author} {\bibfnamefont {D.}~\bibnamefont {Golež}},\
  }\href {https://doi.org/10.21468/SciPostPhys.10.4.091} {\bibfield  {journal}
  {\bibinfo  {journal} {SciPost Phys.}\ }\textbf {\bibinfo {volume} {10}},\
  \bibinfo {pages} {091} (\bibinfo {year} {2021})}\BibitemShut {NoStop}%
\bibitem [{\citenamefont {Karlsson}\ \emph {et~al.}(2021)\citenamefont
  {Karlsson}, \citenamefont {van Leeuwen}, \citenamefont {Pavlyukh},
  \citenamefont {Perfetto},\ and\ \citenamefont
  {Stefanucci}}]{EB_DynamicsGKBA}%
  \BibitemOpen
  \bibfield  {author} {\bibinfo {author} {\bibfnamefont {D.}~\bibnamefont
  {Karlsson}}, \bibinfo {author} {\bibfnamefont {R.}~\bibnamefont {van
  Leeuwen}}, \bibinfo {author} {\bibfnamefont {Y.}~\bibnamefont {Pavlyukh}},
  \bibinfo {author} {\bibfnamefont {E.}~\bibnamefont {Perfetto}},\ and\
  \bibinfo {author} {\bibfnamefont {G.}~\bibnamefont {Stefanucci}},\ }\href
  {https://doi.org/10.1103/PhysRevLett.127.036402} {\bibfield  {journal}
  {\bibinfo  {journal} {Phys. Rev. Lett.}\ }\textbf {\bibinfo {volume} {127}},\
  \bibinfo {pages} {036402} (\bibinfo {year} {2021})}\BibitemShut {NoStop}%
\bibitem [{\citenamefont {Meirinhos}\ \emph {et~al.}(2022)\citenamefont
  {Meirinhos}, \citenamefont {Kajan}, \citenamefont {Kroha},\ and\
  \citenamefont {Bode}}]{meirinhos2022adaptive}%
  \BibitemOpen
  \bibfield  {author} {\bibinfo {author} {\bibfnamefont {F.}~\bibnamefont
  {Meirinhos}}, \bibinfo {author} {\bibfnamefont {M.}~\bibnamefont {Kajan}},
  \bibinfo {author} {\bibfnamefont {J.}~\bibnamefont {Kroha}},\ and\ \bibinfo
  {author} {\bibfnamefont {T.}~\bibnamefont {Bode}},\ }\href
  {https://doi.org/10.21468/SciPostPhysCore.5.2.030} {\bibfield  {journal}
  {\bibinfo  {journal} {SciPost Physics Core}\ }\textbf {\bibinfo {volume}
  {5}},\ \bibinfo {pages} {030} (\bibinfo {year} {2022})}\BibitemShut {NoStop}%
\bibitem [{\citenamefont {Perfetto}\ \emph {et~al.}(2022)\citenamefont
  {Perfetto}, \citenamefont {Pavlyukh},\ and\ \citenamefont
  {Stefanucci}}]{Exciton_GKBA}%
  \BibitemOpen
  \bibfield  {author} {\bibinfo {author} {\bibfnamefont {E.}~\bibnamefont
  {Perfetto}}, \bibinfo {author} {\bibfnamefont {Y.}~\bibnamefont {Pavlyukh}},\
  and\ \bibinfo {author} {\bibfnamefont {G.}~\bibnamefont {Stefanucci}},\
  }\href {https://doi.org/10.1103/PhysRevLett.128.016801} {\bibfield  {journal}
  {\bibinfo  {journal} {Phys. Rev. Lett.}\ }\textbf {\bibinfo {volume} {128}},\
  \bibinfo {pages} {016801} (\bibinfo {year} {2022})}\BibitemShut {NoStop}%
\bibitem [{\citenamefont {Schl\"unzen}\ \emph {et~al.}(2020)\citenamefont
  {Schl\"unzen}, \citenamefont {Joost},\ and\ \citenamefont {Bonitz}}]{G1_G2}%
  \BibitemOpen
  \bibfield  {author} {\bibinfo {author} {\bibfnamefont {N.}~\bibnamefont
  {Schl\"unzen}}, \bibinfo {author} {\bibfnamefont {J.-P.}\ \bibnamefont
  {Joost}},\ and\ \bibinfo {author} {\bibfnamefont {M.}~\bibnamefont
  {Bonitz}},\ }\href {https://doi.org/10.1103/PhysRevLett.124.076601}
  {\bibfield  {journal} {\bibinfo  {journal} {Phys. Rev. Lett.}\ }\textbf
  {\bibinfo {volume} {124}},\ \bibinfo {pages} {076601} (\bibinfo {year}
  {2020})}\BibitemShut {NoStop}%
\bibitem [{\citenamefont {Pavlyukh}\ \emph {et~al.}(2024)\citenamefont
  {Pavlyukh}, \citenamefont {Tuovinen}, \citenamefont {Perfetto},\ and\
  \citenamefont {Stefanucci}}]{Pavlyukh2024}%
  \BibitemOpen
  \bibfield  {author} {\bibinfo {author} {\bibfnamefont {Y.}~\bibnamefont
  {Pavlyukh}}, \bibinfo {author} {\bibfnamefont {R.}~\bibnamefont {Tuovinen}},
  \bibinfo {author} {\bibfnamefont {E.}~\bibnamefont {Perfetto}},\ and\
  \bibinfo {author} {\bibfnamefont {G.}~\bibnamefont {Stefanucci}},\ }\href
  {https://doi.org/https://doi.org/10.1002/pssb.202300504} {\bibfield
  {journal} {\bibinfo  {journal} {physica status solidi (b)}\ }\textbf
  {\bibinfo {volume} {261}},\ \bibinfo {pages} {2300504} (\bibinfo {year}
  {2024})},\ \Eprint
  {https://arxiv.org/abs/https://onlinelibrary.wiley.com/doi/pdf/10.1002/pssb.202300504}
  {https://onlinelibrary.wiley.com/doi/pdf/10.1002/pssb.202300504} \BibitemShut
  {NoStop}%
\bibitem [{\citenamefont {Blommel}\ \emph {et~al.}(2024)\citenamefont
  {Blommel}, \citenamefont {Gardner}, \citenamefont {Woodward},\ and\
  \citenamefont {Gull}}]{Blommel2024}%
  \BibitemOpen
  \bibfield  {author} {\bibinfo {author} {\bibfnamefont {T.}~\bibnamefont
  {Blommel}}, \bibinfo {author} {\bibfnamefont {D.~J.}\ \bibnamefont
  {Gardner}}, \bibinfo {author} {\bibfnamefont {C.~S.}\ \bibnamefont
  {Woodward}},\ and\ \bibinfo {author} {\bibfnamefont {E.}~\bibnamefont
  {Gull}},\ }\href {https://doi.org/10.1103/PhysRevB.110.205134} {\bibfield
  {journal} {\bibinfo  {journal} {Phys. Rev. B}\ }\textbf {\bibinfo {volume}
  {110}},\ \bibinfo {pages} {205134} (\bibinfo {year} {2024})}\BibitemShut
  {NoStop}%
\bibitem [{\citenamefont {Blommel}\ \emph {et~al.}(2025)\citenamefont
  {Blommel}, \citenamefont {Kaye}, \citenamefont {Murakami}, \citenamefont
  {Gull},\ and\ \citenamefont {Gole\ifmmode~\check{z}\else
  \v{z}\fi{}}}]{Blommel2025}%
  \BibitemOpen
  \bibfield  {author} {\bibinfo {author} {\bibfnamefont {T.}~\bibnamefont
  {Blommel}}, \bibinfo {author} {\bibfnamefont {J.}~\bibnamefont {Kaye}},
  \bibinfo {author} {\bibfnamefont {Y.}~\bibnamefont {Murakami}}, \bibinfo
  {author} {\bibfnamefont {E.}~\bibnamefont {Gull}},\ and\ \bibinfo {author}
  {\bibfnamefont {D.}~\bibnamefont {Gole\ifmmode~\check{z}\else \v{z}\fi{}}},\
  }\href {https://doi.org/10.1103/PhysRevB.111.094502} {\bibfield  {journal}
  {\bibinfo  {journal} {Phys. Rev. B}\ }\textbf {\bibinfo {volume} {111}},\
  \bibinfo {pages} {094502} (\bibinfo {year} {2025})}\BibitemShut {NoStop}%
\bibitem [{\citenamefont {Reeves}\ and\ \citenamefont
  {Vlcek}(2024)}]{Reeves2024}%
  \BibitemOpen
  \bibfield  {author} {\bibinfo {author} {\bibfnamefont {C.~C.}\ \bibnamefont
  {Reeves}}\ and\ \bibinfo {author} {\bibfnamefont {V.}~\bibnamefont {Vlcek}},\
  }\href {https://doi.org/10.1103/PhysRevLett.133.226902} {\bibfield  {journal}
  {\bibinfo  {journal} {Phys. Rev. Lett.}\ }\textbf {\bibinfo {volume} {133}},\
  \bibinfo {pages} {226902} (\bibinfo {year} {2024})}\BibitemShut {NoStop}%
\bibitem [{\citenamefont {Yin}\ \emph {et~al.}(2023)\citenamefont {Yin},
  \citenamefont {hao Chan}, \citenamefont {{da Jornada}}, \citenamefont {Qiu},
  \citenamefont {Yang},\ and\ \citenamefont {Louie}}]{DMD1T}%
  \BibitemOpen
  \bibfield  {author} {\bibinfo {author} {\bibfnamefont {J.}~\bibnamefont
  {Yin}}, \bibinfo {author} {\bibfnamefont {Y.}~\bibnamefont {hao Chan}},
  \bibinfo {author} {\bibfnamefont {F.~H.}\ \bibnamefont {{da Jornada}}},
  \bibinfo {author} {\bibfnamefont {D.~Y.}\ \bibnamefont {Qiu}}, \bibinfo
  {author} {\bibfnamefont {C.}~\bibnamefont {Yang}},\ and\ \bibinfo {author}
  {\bibfnamefont {S.~G.}\ \bibnamefont {Louie}},\ }\href
  {https://doi.org/https://doi.org/10.1016/j.jcp.2023.111909} {\bibfield
  {journal} {\bibinfo  {journal} {Journal of Computational Physics}\ }\textbf
  {\bibinfo {volume} {477}},\ \bibinfo {pages} {111909} (\bibinfo {year}
  {2023})}\BibitemShut {NoStop}%
\bibitem [{\citenamefont {Yin}\ \emph {et~al.}(2022)\citenamefont {Yin},
  \citenamefont {hao Chan}, \citenamefont {da~Jornada}, \citenamefont {Qiu},
  \citenamefont {Louie},\ and\ \citenamefont {Yang}}]{DMD2T}%
  \BibitemOpen
  \bibfield  {author} {\bibinfo {author} {\bibfnamefont {J.}~\bibnamefont
  {Yin}}, \bibinfo {author} {\bibfnamefont {Y.}~\bibnamefont {hao Chan}},
  \bibinfo {author} {\bibfnamefont {F.~H.}\ \bibnamefont {da~Jornada}},
  \bibinfo {author} {\bibfnamefont {D.~Y.}\ \bibnamefont {Qiu}}, \bibinfo
  {author} {\bibfnamefont {S.~G.}\ \bibnamefont {Louie}},\ and\ \bibinfo
  {author} {\bibfnamefont {C.}~\bibnamefont {Yang}},\ }\href
  {https://doi.org/https://doi.org/10.1016/j.jocs.2022.101843} {\bibfield
  {journal} {\bibinfo  {journal} {Journal of Computational Science}\ }\textbf
  {\bibinfo {volume} {64}},\ \bibinfo {pages} {101843} (\bibinfo {year}
  {2022})}\BibitemShut {NoStop}%
\bibitem [{\citenamefont {Kemper}\ \emph {et~al.}(2024)\citenamefont {Kemper},
  \citenamefont {Yang},\ and\ \citenamefont {Gull}}]{Extension}%
  \BibitemOpen
  \bibfield  {author} {\bibinfo {author} {\bibfnamefont {A.~F.}\ \bibnamefont
  {Kemper}}, \bibinfo {author} {\bibfnamefont {C.}~\bibnamefont {Yang}},\ and\
  \bibinfo {author} {\bibfnamefont {E.}~\bibnamefont {Gull}},\ }\href
  {https://doi.org/10.1103/PhysRevLett.132.160403} {\bibfield  {journal}
  {\bibinfo  {journal} {Phys. Rev. Lett.}\ }\textbf {\bibinfo {volume} {132}},\
  \bibinfo {pages} {160403} (\bibinfo {year} {2024})}\BibitemShut {NoStop}%
\bibitem [{\citenamefont {Freericks}\ \emph {et~al.}(2009)\citenamefont
  {Freericks}, \citenamefont {Krishnamurthy},\ and\ \citenamefont
  {Pruschke}}]{freericks_theoretical_2009}%
  \BibitemOpen
  \bibfield  {author} {\bibinfo {author} {\bibfnamefont {J.~K.}\ \bibnamefont
  {Freericks}}, \bibinfo {author} {\bibfnamefont {H.~R.}\ \bibnamefont
  {Krishnamurthy}},\ and\ \bibinfo {author} {\bibfnamefont {T.}~\bibnamefont
  {Pruschke}},\ }\href {https://doi.org/10.1103/PhysRevLett.102.136401}
  {\bibfield  {journal} {\bibinfo  {journal} {Phys. Rev. Lett.}\ }\textbf
  {\bibinfo {volume} {102}},\ \bibinfo {pages} {136401} (\bibinfo {year}
  {2009})}\BibitemShut {NoStop}%
\bibitem [{\citenamefont {Perfetto}\ \emph {et~al.}(2016)\citenamefont
  {Perfetto}, \citenamefont {Sangalli}, \citenamefont {Marini},\ and\
  \citenamefont {Stefanucci}}]{perfetto_first-principles_2016}%
  \BibitemOpen
  \bibfield  {author} {\bibinfo {author} {\bibfnamefont {E.}~\bibnamefont
  {Perfetto}}, \bibinfo {author} {\bibfnamefont {D.}~\bibnamefont {Sangalli}},
  \bibinfo {author} {\bibfnamefont {A.}~\bibnamefont {Marini}},\ and\ \bibinfo
  {author} {\bibfnamefont {G.}~\bibnamefont {Stefanucci}},\ }\href
  {https://doi.org/10.1103/PhysRevB.94.245303} {\bibfield  {journal} {\bibinfo
  {journal} {Phys. Rev. B}\ }\textbf {\bibinfo {volume} {94}},\ \bibinfo
  {pages} {245303} (\bibinfo {year} {2016})}\BibitemShut {NoStop}%
\bibitem [{\citenamefont {Latini}\ \emph {et~al.}(2014)\citenamefont {Latini},
  \citenamefont {Perfetto}, \citenamefont {Uimonen}, \citenamefont {van
  Leeuwen},\ and\ \citenamefont {Stefanucci}}]{latini_charge_2014}%
  \BibitemOpen
  \bibfield  {author} {\bibinfo {author} {\bibfnamefont {S.}~\bibnamefont
  {Latini}}, \bibinfo {author} {\bibfnamefont {E.}~\bibnamefont {Perfetto}},
  \bibinfo {author} {\bibfnamefont {A.-M.}\ \bibnamefont {Uimonen}}, \bibinfo
  {author} {\bibfnamefont {R.}~\bibnamefont {van Leeuwen}},\ and\ \bibinfo
  {author} {\bibfnamefont {G.}~\bibnamefont {Stefanucci}},\ }\href
  {https://doi.org/10.1103/PhysRevB.89.075306} {\bibfield  {journal} {\bibinfo
  {journal} {Phys. Rev. B}\ }\textbf {\bibinfo {volume} {89}},\ \bibinfo
  {pages} {075306} (\bibinfo {year} {2014})}\BibitemShut {NoStop}%
\bibitem [{\citenamefont {Tuovinen}\ \emph {et~al.}(2023)\citenamefont
  {Tuovinen}, \citenamefont {Pavlyukh}, \citenamefont {Perfetto},\ and\
  \citenamefont {Stefanucci}}]{tuovinen_time-linear_2023}%
  \BibitemOpen
  \bibfield  {author} {\bibinfo {author} {\bibfnamefont {R.}~\bibnamefont
  {Tuovinen}}, \bibinfo {author} {\bibfnamefont {Y.}~\bibnamefont {Pavlyukh}},
  \bibinfo {author} {\bibfnamefont {E.}~\bibnamefont {Perfetto}},\ and\
  \bibinfo {author} {\bibfnamefont {G.}~\bibnamefont {Stefanucci}},\ }\href
  {https://link.aps.org/doi/10.1103/PhysRevLett.130.246301} {\bibfield
  {journal} {\bibinfo  {journal} {Phys. Rev. Lett.}\ }\textbf {\bibinfo
  {volume} {130}},\ \bibinfo {pages} {246301} (\bibinfo {year}
  {2023})}\BibitemShut {NoStop}%
\bibitem [{\citenamefont {Hedayat}\ \emph {et~al.}(2021)\citenamefont
  {Hedayat}, \citenamefont {Ceraso}, \citenamefont {Soavi}, \citenamefont
  {Akhavan}, \citenamefont {Cadore}, \citenamefont {Dallera}, \citenamefont
  {Cerullo}, \citenamefont {Ferrari},\ and\ \citenamefont
  {Carpene}}]{Hedayat_2021}%
  \BibitemOpen
  \bibfield  {author} {\bibinfo {author} {\bibfnamefont {H.}~\bibnamefont
  {Hedayat}}, \bibinfo {author} {\bibfnamefont {A.}~\bibnamefont {Ceraso}},
  \bibinfo {author} {\bibfnamefont {G.}~\bibnamefont {Soavi}}, \bibinfo
  {author} {\bibfnamefont {S.}~\bibnamefont {Akhavan}}, \bibinfo {author}
  {\bibfnamefont {A.}~\bibnamefont {Cadore}}, \bibinfo {author} {\bibfnamefont
  {C.}~\bibnamefont {Dallera}}, \bibinfo {author} {\bibfnamefont
  {G.}~\bibnamefont {Cerullo}}, \bibinfo {author} {\bibfnamefont {A.~C.}\
  \bibnamefont {Ferrari}},\ and\ \bibinfo {author} {\bibfnamefont
  {E.}~\bibnamefont {Carpene}},\ }\href
  {https://doi.org/10.1088/2053-1583/abd89a} {\bibfield  {journal} {\bibinfo
  {journal} {2D Materials}\ }\textbf {\bibinfo {volume} {8}},\ \bibinfo {pages}
  {025020} (\bibinfo {year} {2021})}\BibitemShut {NoStop}%
\bibitem [{\citenamefont {Spataru}\ \emph {et~al.}(2004)\citenamefont
  {Spataru}, \citenamefont {Benedict},\ and\ \citenamefont
  {Louie}}]{Spataru_2004}%
  \BibitemOpen
  \bibfield  {author} {\bibinfo {author} {\bibfnamefont {C.~D.}\ \bibnamefont
  {Spataru}}, \bibinfo {author} {\bibfnamefont {L.~X.}\ \bibnamefont
  {Benedict}},\ and\ \bibinfo {author} {\bibfnamefont {S.~G.}\ \bibnamefont
  {Louie}},\ }\href {https://doi.org/10.1103/PhysRevB.69.205204} {\bibfield
  {journal} {\bibinfo  {journal} {Phys. Rev. B}\ }\textbf {\bibinfo {volume}
  {69}},\ \bibinfo {pages} {205204} (\bibinfo {year} {2004})}\BibitemShut
  {NoStop}%
\bibitem [{\citenamefont {Callan}\ \emph {et~al.}(2000)\citenamefont {Callan},
  \citenamefont {Kim}, \citenamefont {Huang},\ and\ \citenamefont
  {Mazur}}]{CALLAN2000167}%
  \BibitemOpen
  \bibfield  {author} {\bibinfo {author} {\bibfnamefont {J.}~\bibnamefont
  {Callan}}, \bibinfo {author} {\bibfnamefont {A.~M.-T.}\ \bibnamefont {Kim}},
  \bibinfo {author} {\bibfnamefont {L.}~\bibnamefont {Huang}},\ and\ \bibinfo
  {author} {\bibfnamefont {E.}~\bibnamefont {Mazur}},\ }\href
  {https://doi.org/https://doi.org/10.1016/S0301-0104(99)00301-8} {\bibfield
  {journal} {\bibinfo  {journal} {Chemical Physics}\ }\textbf {\bibinfo
  {volume} {251}},\ \bibinfo {pages} {167} (\bibinfo {year}
  {2000})}\BibitemShut {NoStop}%
\end{thebibliography}%
\bibliographystyle{apsrev4-2}

\end{document}